\documentclass[aps,preprint,showpacs,preprintnumbers,amsmath,amssymb]{revtex4}
\usepackage{amsmath,mathrsfs,amsbsy,color,graphicx,bm,amsthm,amsfonts}
\usepackage{units}
\usepackage{bbm}
\usepackage{times}
\usepackage{dcolumn}
\usepackage{mathrsfs}
\usepackage{amsmath,amssymb,epsfig}
\usepackage{amsmath}
\newcommand{\udots}{\mathinner{\mskip1mu\raise1pt\vbox{\kern7pt\hbox{.}}
\mskip2mu\raise4pt\hbox{.}\mskip2mu\raise7pt\hbox{.}\mskip1mu}}
\begin{document}
\title{Multiqubit coherence of mixed states near event horizon }
\author{Wen-Mei Li$^1$,  Jianbo Lu$^1$\footnote{ lvjianbo819@163.com (corresponding author)}, Shu-Min Wu$^1$\footnote{smwu@lnnu.edu.cn (corresponding author)} }
\affiliation{$^1$ Department of Physics, Liaoning Normal University, Dalian 116029, China}


\begin{abstract}
We investigate the coherence of mixed Greenberger-Horne-Zeilinger (GHZ) and W states for bosonic and fermionic fields when a subset of  $n$ ($n<N$) qubits experiences Hawking radiation near a Schwarzschild black hole. Analytical expressions are derived for the coherence of mixed N-qubit systems, including both the physically accessible and inaccessible parts in curved spacetime. The results show that the mixed W state maintains its coherence more effectively than the GHZ state as the Hawking temperature increases, even though its entanglement is weaker. As the number of qubits grows, W-state coherence becomes increasingly resistant to gravitational decoherence. Furthermore, fermionic fields preserve stronger entanglement, while bosonic fields retain higher coherence, highlighting a clear contrast between different particle statistics. These findings demonstrate how the Schwarzschild spacetime reshapes the balance between quantum coherence and entanglement, offering guidance for future relativistic quantum information applications.
\end{abstract}

\vspace*{0.5cm}
\maketitle
\section{Introduction}
Quantum coherence, arising from the phenomenon of quantum superposition, lies at the very heart of quantum mechanics. It is not only a fundamental prerequisite for quantum entanglement and various other forms of quantum correlation, but it also serves as a critical resource for quantum computation and information processing \cite{L111,L112,L113}. Beyond its theoretical importance, quantum coherence has profound practical implications, playing an indispensable role in fields such as nanoscale thermodynamics \cite{L7,L8,L9}, energy transport in biological systems \cite{L6}, quantum cryptography \cite{L3,L4,L5}, and quantum metrology \cite{L1,L2}.
As coherence extends from single-particle to multipartite systems, it enables advanced applications like coherence localization and incoherent teleportation \cite{L10,L11}. The quantification of quantum coherence, first proposed by Baumgratz $et$ $al.$ \cite{L12} within the context of resource theory, introducing two key measures: the relative entropy of coherence (REC) and the $l_{1}$ norm of coherence \cite{L12,L13,L14,L15}.  While REC offers a deep understanding of coherence structure, the $l_{1}$ norm provides distinct advantages due to its analytical simplicity, ease of calculation, and clearer physical interpretation, especially in multipartite systems where numerical methods may be complex. This simplicity makes the $l_{1}$ norm an invaluable tool for both theoretical exploration and practical quantum technologies.

Relativistic quantum information \cite{QFK3,QFK4,QFK5,QFK6,R3,R1,R2,WDD4,RM1,RM3,RM5,B31,R15,R5,R6,R7,R8,R9,BJ1,BJ2,BJ3,R10,R11,R12,R13,R14,R16,R17,R18,RM2,RM4,RM6,RM7,RM8,RM9,RM10,RM11,RM12,QQRM12,WDD1,WDD2,WDD3,BJ4,R4,WDD5,QQQQR8,qert1,RTYR4,L36,QFK1,QFK2} integrates elements of information theory, quantum optics, quantum field theory, and general relativity. It is crucial to comprehend quantum information within the framework of relativity, as gravitational effects on quantum systems have become a non-negligible critical factor in feasible long-distance experimental scenarios. Traditionally, relativistic quantum information focused on the effects of gravity on bipartite or tripartite pure-state systems \cite{R3,R1,R2,R4,RM1,RM3,RM5,B31,R5,R6,R7,R8,R9,BJ1,BJ2,BJ3,R10,R11,R12,R13,R14,R15,R16,R17,R18,RM2,RM4,RM6,RM7,RM8,RM9,RM10,RM11,RM12,QQRM12,WDD1,WDD2,WDD3,BJ4,WDD4,WDD5,QQQQR8,qert1,RTYR4,L36}. However, in realistic quantum systems, pure states exhibit inherent susceptibility to environmental decoherence, inevitably transitioning into statistical mixtures. Whereas under laboratory conditions, preserving quantum purity demands sophisticated environmental decoherence mitigation protocols and stringent isolation engineering \cite{M11,M12,M13}. This paper introduces global noise to pure states, transforming them into mixed states. Moreover, as quantum information tasks grow increasingly complex, conventional bipartite or tripartite quantum resources face limitations, prompting the need to explore N-partite quantum resources to meet advanced demands. Landmark advancements include: Deng et al.'s groundbreaking observation of 255 photon-click events through pseudo-photon-number-resolving detection in novel Gaussian boson sampling experiments \cite{Q13}; and Google Quantum AI's collaborative achievement in stabilizing surface code memory below theoretical thresholds via engineered 105-qubit nonseparable systems within the Willow architecture \cite{Q14}. Based on the above two practical reasons, it is necessary to study bosonic and fermionic coherence of N-qubit mixed states in the background of the Schwarzschild black hole. It is one of the motivations for studying the N-qubit coherence of the mixed GHZ and W states in the Schwarzschild black hole. Due to computational complexity, the redistribution of quantum coherence in N-qubit mixed states under the influence of Hawking radiation remains insufficiently understood. This is another motivation for our research to understand the laws of quantum coherence for N-qubit mixed states shuttling inside and outside the event horizon.

This work investigates N-qubit coherence of the mixed GHZ and W states for bosonic and fermionic fields configurations in the Schwarzschild spacetime. We begin by considering $N$ observers who initially share the mixed GHZ and W states in asymptotically flat region. The scenario involves a gravitational setting in which $N-n$ observers remain inertial in flat spacetime, while $n$ observers hover near the event horizon. We will derive comprehensive analytical expressions to quantify multipartite coherence in mixed states, accounting for both physically accessible and inaccessible components in curved spacetime. We will compare the coherence properties of the mixed GHZ and W states and analyze the distinctions between quantum coherence and entanglement for bosonic and fermionic fields in the Schwarzschild background. Furthermore, we will investigate how the Hawking effect redistributes multipartite coherence and reveal the underlying dynamics governing the flow of multiqubit quantum information across the event horizon. From the perspective of relativistic quantum information tasks, our findings uncover how the interplay between particle statistics and  Schwarzschild black hole critically shapes the trade-off between coherence and entanglement, offering valuable insights for optimizing quantum information processing in strong gravitational fields.

The structure of the paper is as follows. In Sec. II, we introduce the quantization of bosonic and fermionic fields in Schwarzschild spacetime.  In Sec. III, we discuss multiqubit coherence of mixed states for bosonic and fermionic fields in the Schwarzschild black hole. The last section is devoted to the summary.

\section{The quantization of field in Schwarzschild spacetime \label{GSCDGE}}
The metric of a Schwarzschild black hole \cite{B31} can be written as
\begin{eqnarray}\label{S1}
ds^{2}=-\bigg(1-\frac{2M}{r}\bigg)dt^{2}+\bigg(1-\frac{2M}{r}\bigg)^{-1}dr^{2}+r^{2}(d\theta^{2}+\sin^{2}\theta d\varphi^{2}),
\end{eqnarray}
where parameters $M$ and $r$ represent the mass and radius of the Schwarzschild black hole, respectively. For convenience of calculation, we let $h=G=c=k=1$ in this paper.
\subsection{Bosonic field}
The dynamics of a massless scalar field satisfies the Klein-Gordon equation \cite{R6,M5}
\begin{eqnarray}\label{S2}
\frac{1}{\sqrt{-g}}\frac{\partial}{\partial x^{\mu}}(\sqrt{-g}g^{\mu\nu}\frac{\partial\Psi}{\partial x^{\nu}})=0.
\end{eqnarray}
The normal mode solution can be written as
\begin{eqnarray}\label{S3}
\Psi_{\omega lm}=\frac{1}{R(r)}\chi_{\omega l}(r)Y_{lm}(\theta,\varphi)e^{-i\omega t},
\end{eqnarray}
and we can easily obtain the radial equation as
\begin{eqnarray}\label{S4}
\frac{d^{2}\chi_{\omega l}}{dr^{2}_{\ast}}+[\omega^{2}-V(r)]\chi_{\omega l}=0.
\end{eqnarray}
Solving Eq.(\ref{S4}), one can get the incoming wave function in the spacetime manifold which is analytic everywhere
\begin{eqnarray}\label{S5}
\Psi_{in,\omega lm}=e^{-i\omega \nu}Y_{lm}(\theta,\varphi),
\end{eqnarray}
and the outgoing wave functions for the outside and inside regions of the event horizon are given by
\begin{eqnarray}\label{S6}
\Psi_{out,\omega lm}(r>r_{+})=e^{-i\omega \mu}Y_{lm}(\theta,\varphi),
\end{eqnarray}
\begin{eqnarray}\label{S7}
\Psi_{out,\omega lm}(r<r_{+})=e^{i\omega \mu}Y_{lm}(\theta,\varphi),
\end{eqnarray}
where $\nu=t+r_{\ast}$ and $\mu=t-r_{\ast}$.
Eqs.(\ref{S6}) and (\ref{S7}) are separately analytic
outside and inside the event horizon, respectively, thus constituting a completely orthogonal family. In the process of second quantization of the scalar field outside the event horizon, the field  is expanded as
\begin{eqnarray}\label{S8}
\Phi_{out}&=&\sum_{lm}\int d\omega[b_{in,\omega lm}\Psi_{in,\omega lm}(r<r_{+})
+b_{in,\omega lm}^{\dag}\Psi_{in,\omega lm}^{\ast}(r<r_{+})\notag\\
&+&b_{out,\omega lm}\Psi_{out,\omega lm}(r>r_{+})+b_{out,\omega lm}^{\dag}\Psi_{out,\omega lm}^{\ast}(r>r_{+})],
\end{eqnarray}
where $b_{in,\omega lm}$ and $b_{in,\omega lm}^{\dagger}$ serve as the annihilation and creation operators for the vacuum state within the interior of the Schwarzschild black hole, while $b_{out,\omega lm}$ and $b_{out,\omega lm}^{\dagger}$ correspond to the exterior region. The Fock vacuum states satisfy
\begin{eqnarray}\label{S9}
b_{in,\omega lm}|0\rangle_{in}=b_{out,\omega lm}|0\rangle_{out}=0.
\end{eqnarray}

One introduces the generalized light-like Kruskal coordinates
\begin{eqnarray}\label{S10}
&&U=4Me^{-\frac{\mu}{4M}},V=4Me^{\frac{\nu}{4M}},\mathrm{if}\;r<r_{+};\notag\\
&&U=-4Me^{-\frac{\mu}{4M}},V=4Me^{\frac{\nu}{4M}},\mathrm{if}\;r>r_{+}.
\end{eqnarray}
Therefore, we can rewrite the Schwarzschild modes as
\begin{eqnarray}\label{S11}
\Psi_{out,\omega lm}(r<r_{+})=e^{-4i\omega M\ln[-\frac{U}{4M}]}Y_{lm}(\theta,\varphi),
\end{eqnarray}
\begin{eqnarray}\label{S12}
\Psi_{out,\omega lm}(r>r_{+})=e^{4i\omega M\ln[\frac{U}{4M}]}Y_{lm}(\theta,\varphi).
\end{eqnarray}
Following the method proposed by Ruffini and Damour \cite{M7,M8}, one constructs a complete orthonormal basis of outgoing modes
\begin{eqnarray}\label{S13}
\Psi_{I,\omega lm}=e^{2\pi\omega M}\Psi_{out,\omega lm}(r>r_{+})
+e^{-2\pi\omega M}\Psi_{out,\omega lm}^{\ast}(r<r_{+}),
\end{eqnarray}
\begin{eqnarray}\label{S14}
\Psi_{II,\omega lm}=e^{-2\pi\omega M}\Psi_{out,\omega lm}^{\ast}(r>r_{+})
+e^{2\pi\omega M}\Psi_{out,\omega lm}(r<r_{+}).
\end{eqnarray}
Thus, we can also quantize the scalar field $\Phi_{out}$ in terms of $\Psi_{I,\omega lm}$ and $\Psi_{II,\omega lm}$ in the Kruskal spacetime as
\begin{eqnarray}\label{S15}
\Phi_{out}&=&\sum_{lm} \int d\omega[2\sin(4\pi\omega M)]^{-1/2}[a_{out,\omega lm}\Psi_{I,\omega lm}\notag\\
&+&a_{out,\omega lm}^{\dagger}\Psi_{I,\omega lm}^{\ast}+a_{in,\omega lm}\Psi_{II,\omega lm}+a_{in,\omega lm}^{\dagger}\Psi_{II,\omega lm}^{\ast}],
\end{eqnarray}
where  $a_{out,\omega lm}$ defines the annihilation operator associated with the Kruskal vacuum state
\begin{eqnarray}\label{S16}
a_{out,\omega lm}|0\rangle_{K}=0.
\end{eqnarray}
According to Eqs.(\ref{S8}) and (\ref{S15}), we obtain the Bogoliubov relation for the bosonic annihilation and creation operators in the Schwarzschild black hole and Kruskal spacetime as
\begin{eqnarray}\label{S17}
a_{out,\omega lm}^{B}=\frac{b_{out,\omega lm}^{B}}{\sqrt{1-e^{-\omega/T}}}-\frac{b_{in,\omega lm}^{B,\dag}}{\sqrt{e^{\omega/T}-1}},
\end{eqnarray}
\begin{eqnarray}\label{S18}
a_{out,\omega lm}^{B,\dag}=\frac{b_{out,\omega lm}^{B,\dag}}{\sqrt{1-e^{-\omega/T}}}
-\frac{b_{in,\omega lm}^{B}}{\sqrt{e^{\omega/T}-1}},
\end{eqnarray}
where $T=\frac{1}{8\pi M}$ represents Hawking temperature \cite{R6,MM9}.

By normalizing the state vector, the Kruskal vacuum state of the bosonic field  in the Schwarzschild black hole can be written as a maximally entangled two-mode squeezed state
\begin{eqnarray}\label{S19}
|0\rangle_{K}^{B}=\sqrt{1-e^{-\omega/T}}\sum^{\infty}_{n=0} e^{-n\omega/2T}|n\rangle_{out}^{B}|n\rangle_{in}^{B},
\end{eqnarray}
and the first excited state of bosonic field reads
\begin{eqnarray}\label{S20}
|1\rangle_{K}^{B}=a_{out,\omega lm}^{\dagger}|0\rangle _{K}^{B}=
(1-e^{-\omega/T})\sum^{\infty}_{n=0}\sqrt{n+1} e^{-n\omega/2T}|n+1\rangle_{out}^{B}|n\rangle_{in}^{B},
\end{eqnarray}
where $B$ represents the bosonic field, $\{|n\rangle_{out}\}$ and $\{|n\rangle_{in}\}$ represent orthonormal number states outside and inside the event horizon, respectively \cite{RTYR4}.

The Schwarzschild observer hovers outside the event horizon and detects a Hawking radiation spectrum given by
\begin{eqnarray}\label{S21}
N_{\omega}^{B}=\frac{1}{e^{\frac{\omega}{T}}-1}.
\end{eqnarray}
From the Schwarzschild observer’s viewpoint, this equation indicates that the Kruskal vacuum state appears as a thermal bath populated by a certain number of bosons $N_{\omega}^{B}$. In other words, the observer perceives a thermal distribution characteristic of Bose-Einstein statistics.

\subsection{Fermionic field}
Conducting a computation analogous to that for bosonic field, one obtains the Bogoliubov transformation between the Kruskal and Schwarzschild operators for fermionic field as \cite{R18}
\begin{eqnarray}\label{S22}
a_{out,\omega lm}^{F}=\frac{a_{out,\omega lm}^{F}}{\sqrt{e^{-\omega/T}+1}}-\frac{b_{in,\omega lm}^{F,\dag}}{\sqrt{e^{\omega/T}+1}},
\end{eqnarray}
\begin{eqnarray}\label{S23}
a_{out,\omega lm}^{F,\dag}=\frac{a_{out,\omega lm}^{F,\dag}}{\sqrt{e^{-\omega/T}+1}}
-\frac{b_{in,\omega lm}^{F}}{\sqrt{e^{\omega/T}+1}}.
\end{eqnarray}
After properly normalizing the state vector, the Kruskal vacuum and the first excited state of fermionic field in curved spacetime can be expressed as
\begin{eqnarray}\label{S24}
|0\rangle_{K}^{F}=\frac{1}{\sqrt{e^{-\omega/T}+1}}|0\rangle_{out}^{F}|0\rangle_{in}^{F}+\frac{1}{\sqrt{e^{\omega/T}+1}}|1\rangle_{out}^{F}|1\rangle_{in}^{F},
\end{eqnarray}
and
\begin{eqnarray}\label{S25}
|1\rangle_{K}^{F}=|1\rangle_{out}^{F}|0\rangle_{in}^{F},
\end{eqnarray}
where $F$ represents fermionic field. By analogy with bosonic case, one can infer the Hawking radiation spectrum for fermionic field as
\begin{eqnarray}\label{S26}
N_{\omega}^{F}=\frac{1}{e^{\frac{\omega}{T}}+1}.
\end{eqnarray}
This result shows that an observer outside the event horizon perceives a thermal distribution of particles obeying Fermi-Dirac statistics. Comparing Eqs.(\ref{S21}) and (\ref{S26}), we see that bosonic and fermionic fields yield distinct statistical distributions, leading to influences of different gravitational effects on quantum Resources in Schwarzschild spacetime. These differences influence both the coherence and the entanglement properties of the respective quantum fields.

\section{Multiqubit coherence of mixed states in Schwarzschild black hole  \label{GSCDGE}}
\begin{figure}[htbp]
  \centering
  \includegraphics[width=0.5\textwidth]{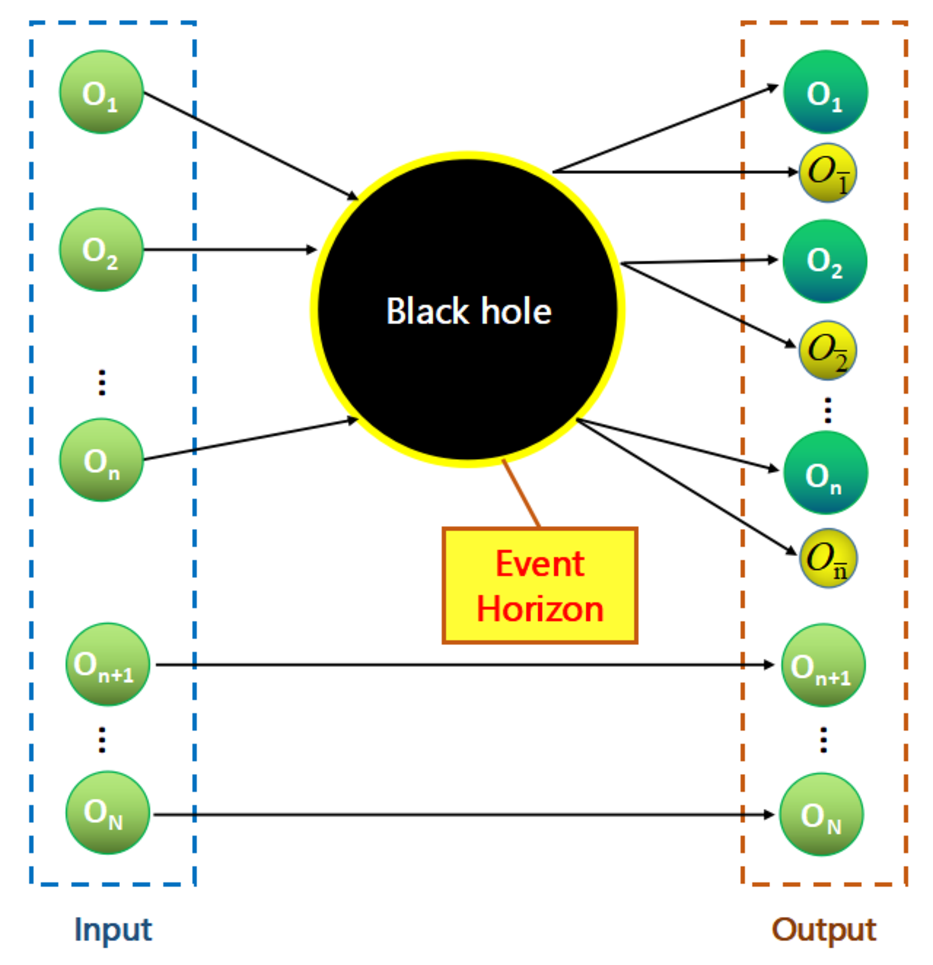}
  \caption{ Schematic diagram of our physical model with $N-n$ observers in a flat region, and $n$ observers near the event horizon of a Schwarzschild black hole. }
  \label{Fig8}
\end{figure}

In realistic physical systems, pure quantum states are highly vulnerable to environmental decoherence and often evolve into mixed states. Maintaining quantum purity in laboratory conditions demands stringent isolation and precise control. To model this inevitable decoherence, we introduce global depolarizing noise to all pure states in this study. Specifically, the mixed N-qubit GHZ state is defined as
\begin{eqnarray}\label{S29}
\mathcal{\tilde{\rho}}^{B/F}(GHZ)=\frac{p}{2^{N}}I_{d}+\big(1-p\big)|GHZ_{123 \cdots N}\rangle\langle GHZ_{123 \cdots N}|,
\end{eqnarray}
where the pure N-qubit GHZ state ($B/F$ denoting boson/ fermion) is defined as
\begin{eqnarray}\label{S27}
|GHZ^{B/F}_{123\ldots N}\rangle=\frac{1}{\sqrt{2}}(|00\ldots00\rangle+|11\ldots11\rangle).
\end{eqnarray}
Here,  $I_{d}$ is the $2^{N}\times2^{N}$ identity matrix, and $p\in [0,1]$ serves as the noise mixing parameter. The parameter $p$ quantifies the degree of depolarization: $p=0$ corresponds to the ideal pure GHZ state, while $p=1$ yields the maximally mixed state. Similarly, the mixed  N-qubit W state is constructed as
\begin{eqnarray}\label{S30}
\mathcal{\tilde{\rho}}^{B/F}(W)=\frac{p}{2^{N}}I_{d}+\big(1-p\big)|W_{123 \cdots N}\rangle\langle W_{123 \cdots N}|,
\end{eqnarray}
with the pure N-qubit W state given by
\begin{eqnarray}\label{S28}
|W^{B/F}_{123\ldots N}\rangle&=&\frac{1}{\sqrt{N}}(|10\ldots 00\rangle+|01\ldots 00\rangle+\cdots+|00\ldots 01\rangle),
\end{eqnarray}
at the same point in the asymptotically flat region of the Schwarzschild black hole.
After sharing their own qubit, $n$ $(1\leq n\leq N)$ observers hover near the event horizon of the Schwarzschild black hole, while the remaining $N-n$ observers stay stationary in an asymptotically flat region (see Fig.\ref{Fig8}).

\begin{figure}[htbp]
  \centering
  \includegraphics[width=0.6\textwidth]{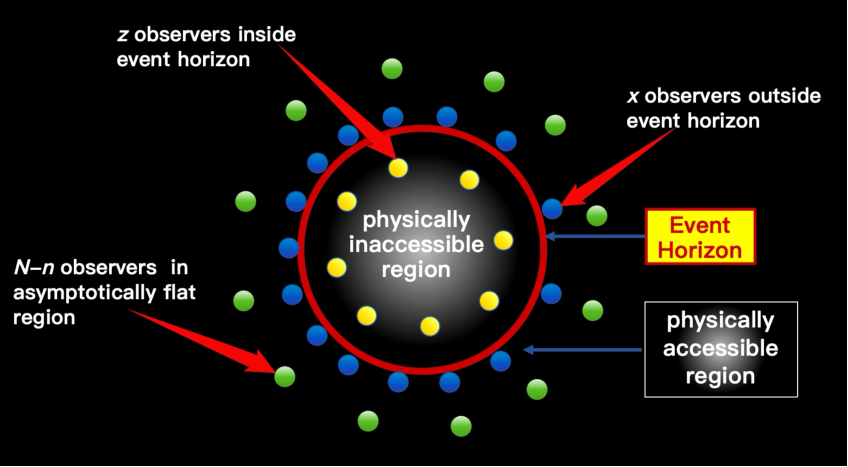}
  \caption{Schematic diagram illustrating the numbers of physically accessible and inaccessible modes in the Schwarzschild black hole.}
  \label{Fig9}
\end{figure}

This study examines the coherence properties of mixed $N$-qubit states in the Schwarzschild black hole spacetime, where distinguishing between the physically accessible region and the physically inaccessible region is crucial for understanding the causal structure of field modes. The physically accessible region lies outside the event horizon, where field modes remain in causal contact with external observers and can be locally measured. In contrast, the physically inaccessible region is confined within the event horizon, where information is causally disconnected from the exterior. To provide an intuitive visualization of the relationship between spacetime geometry and quantum accessibility, Fig.\ref{Fig9} presents a schematic representation highlighting the event horizon and delineating the accessible and inaccessible regions. The figure illustrates the numbers of physically accessible and inaccessible modes in the Schwarzschild black hole, providing a geometric reference for decomposing accessible and inaccessible coherence.

It is worth noting that the global depolarizing noise model adopted in this work assumes a uniform decoherence process acting identically on all qubits. This simplification enables analytical treatment of the N-qubit mixed states and effectively describes collective noise sources that symmetrically affect the entire system. In more realistic scenarios, however, the noise experienced by each observer may vary with their gravitational potential or proximity to the event horizon. Extending the present framework to include local or observer-dependent decoherence channels would provide a more complete picture of open-system dynamics in curved spacetime and may uncover asymmetries in coherence redistribution between near-horizon and distant observers.

\subsection{Multiqubit coherence of mixed GHZ and W states for bosonic field}
 Since the diagonal component of the mixed GHZ state density operator does not contribute to coherence calculations, we focus exclusively on its non-diagonal part. Therefore, we firstly focus on the pure GHZ state in the Schwarzschild black hole.
We use Schwarzschild modes of bosonic field defined in Eqs.(\ref{S19}) and (\ref{S20}) to reformulate Eq.(\ref{S27})
\begin{eqnarray}\label{S31}
GHZ^{B}_{123\ldots N+n}&=&\frac{1}{\sqrt{2}}\bigg\{(1-e^{-\frac{\omega}{T}})^{\frac{n}{2}}\overbrace{(|0\rangle_{N}|0\rangle_{N-1}\cdots|0\rangle_{n+1})}^{|\bar{0}\rangle}\notag\\
&&\bigg[\big[\prod^{n}_{i=1}\sum^{\infty}_{m_{i}=0}e^{\frac{-m_{i}\omega}{2T}}\big]\big[\bigotimes^{n}_{i=1}|m_{i}\rangle_{out}|m_{i}\rangle_{in}\big]\bigg]\notag\\
&&+(1-e^{-\frac{\omega}{T}})^{n}\overbrace{(|1\rangle_{N}|1\rangle_{N-1}\cdots|1\rangle_{n+1})}^{|\bar{1}\rangle}\notag\\
&&\bigg[\big[\prod^{n}_{i=1}\sum^{\infty}_{m_{i}=0}e^{\frac{-m_{i}\omega}{2T}}\sqrt{m+1_{i}}\big]\big[\bigotimes^{n}_{i=1}|m+1_{i}\rangle_{out}|m_{i}\rangle_{in}\big]\bigg]\bigg\},
\end{eqnarray}
where $|\bar{0}\rangle=|0\rangle_{n+1}|0\rangle_{n+2}\cdots|0\rangle_{N}$ and $|\bar{1}\rangle=|1\rangle_{n+1}|1\rangle_{n+2}\cdots|1\rangle_{N}$. Then,
to examine the physically accessible and inaccessible quantum coherences of the mixed GHZ state, we define a composite system $\mathcal{\rho}(GHZ_{N}^{B})$, comprising $N-n$ modes in the asymptotically flat region, $x$ accessible modes outside the event horizon, and $z$ inaccessible modes inside the event horizon.
These satisfy the relation $x+z=n$ , where $n$ is the number of particles affected by the Hawking effect. Accordingly,
the total density operator of the mixed N-qubit GHZ state in Eq.(\ref{S29}) decomposes as
\begin{eqnarray}\label{S32}
\mathcal{\rho}(GHZ_{N}^{B})=\mathcal{\rho}^{GHZ_{N}^{B}}_{diag}+\mathcal{\rho}^{GHZ_{N}^{B}}_{non-diag},
\end{eqnarray}
where $\mathcal{\rho}^{GHZ_{N}^{B}}_{diag}$ is the diagonal and non-diagonal parts, and the non-diagonal density operator  takes the form
\begin{eqnarray}\label{S33}
\mathcal{\rho}^{GHZ_{N}^{B}}_{non-diag}&=&\frac{\big(1-p\big)}{2}\bigg\{\big[(1-e^{\frac{-\omega}{T}})\big]^{\frac{3n}{2}}\prod_{i=1}^{x}\big[\sum^{\infty}_{m_{i}=0}\alpha_{m_i}\beta_{m_i}\big]\prod_{j=1}^{z}\big[\sum^{\infty}_{m_{j}=0}\beta_{m_j}\gamma_{m_j}\big]|\overline{0}\rangle\langle\overline{1}|\notag\\
&\times&\big[\bigotimes^{x}_{i=1}|m_{i}\rangle _{out}\langle m+1_{i}|\big]\big[\bigotimes^{z}_{j=1}|m+1_{j}\rangle _{in}\langle m_{j}|\big]+\big[(1-e^{\frac{-\omega}{T}})\big]^{\frac{3n}{2}}\notag\\
&\times&\prod_{i=1}^{x}\big[\sum^{\infty}_{m_{i}=0}\alpha_{m_i}\beta_{m_i}\big]\prod_{j=1}^{z}\big[\sum^{\infty}_{m_{j}=0}\beta_{m_j}\gamma_{m_j}\big]\overline{1}\rangle\langle\overline{0}|\big[\bigotimes^{x}_{i=1}|m+1_{i}\rangle _{out}\langle m_{i}|\big]\notag\\
&\times&\big[\bigotimes^{z}_{j=1}|m_{j}\rangle _{in}\langle m+1_{j}|\big]\bigg\},
\end{eqnarray}
with the coefficients
\begin{equation}\nonumber
\begin{aligned}
\alpha_{m_i}=e^{\frac{-m_{i}\omega}{2T}}, \notag\\
\beta_{m_i}=\sqrt{m+1_{i}}e^{\frac{-m_{i}\omega}{2T}}, \notag\\
\gamma_{m_i}=e^{\frac{-(m_{i}+1)\omega}{2T}}.
\end{aligned}
\end{equation}

In this study, we quantify quantum coherence in  Schwarzschild spacetime using the $l_{1}$ norm measure. The $l_{1}$ norm of coherence is defined as the sum of the absolute values of all off-diagonal elements in the system’s density matrix
\begin{eqnarray}\label{S34}
C(\rho)=\sum_{i\neq j}|\rho_{i,j}|.
\end{eqnarray}
Using Eq.(\ref{S34}), we derive the analytical expression for the N-qubit coherence of the mixed GHZ state associated with a bosonic field in curved spacetime as
\begin{eqnarray}\label{S35}
C^{B}(GHZ)&=&\big(1-p\big)\bigg[\big(1-e^{\frac{-\omega}{T}}\big)^{\frac{3}{2}}\sum^{\infty}_{i=0}\sqrt{i+1}\big(e^{\frac{-m\omega}{2T}}\big)^{2i}\bigg]^{x}\notag\\
&\times&\bigg[\big(1-e^{\frac{-\omega}{T}}\big)^{\frac{3}{2}}\sum^{\infty}_{i=0}\sqrt{i+1}\big(e^{\frac{-m\omega}{2T}}\big)^{2i+1}\bigg]^{z}.
\end{eqnarray}

\begin{table}[htbp]
  \centering
  \includegraphics[width=0.6\textwidth]{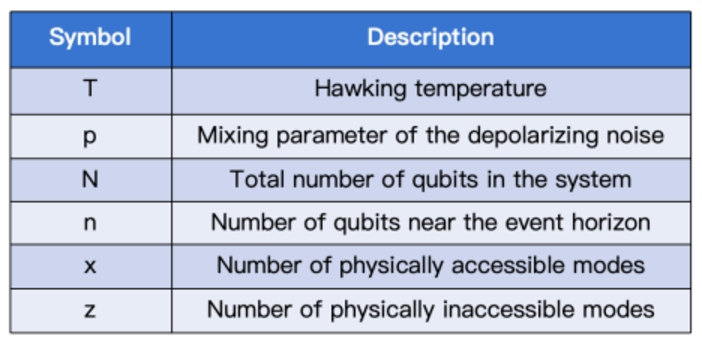}
  \caption{The symbol parameter table.}
  \label{Table 1}
\end{table}

We provide a summary table of parameters (see Table \ref{Table 1}) to clarify the physical meanings and interrelations of the key quantities used in the paper. The table provides a systematic summary of all parameters, including their symbols and definitions, to clarify the notation used in the subsequent analysis. The  parameters include the Hawking temperature $T$, the noise mixing parameter $p$, the total number of qubits in the system $N$, and the number of qubits located near the event horizon $n$. The parameter $n$ can be further decomposed into the number of physically accessible modes $x$ and physically inaccessible modes $z$, satisfying the relation $x + z = n$.

Analogous to the GHZ state formulation for bosonic fields, the N-qubit W state in Schwarzschild spacetime can be expressed in terms of Schwarzschild modes as
\begin{eqnarray}\label{S36}
W^{B}_{123\ldots N+n}
&=&\frac{1}{\sqrt{N}}\bigg\{(1-e^{-\frac{\omega}{T})^{\frac{n}{2}}}|1\rangle_{_{N}}|0\rangle_{_{N-1}}\cdots|0\rangle_{_{n+1}}\big[\prod^{n}_{i=1}\sum^{\infty}_{m_{i}=0}\alpha_{i}\big]\notag\\
&\times&\big[\bigotimes^{n}_{i=1}|m_{i}\rangle_{out}|m_{i}\rangle_{in}\big]+(1-e^{-\frac{\omega}{T}})^{\frac{n}{2}}|0\rangle_{_{N}}|1\rangle_{_{N-1}}\cdots|0\rangle_{_{n+1}}\big[\prod^{n}_{i=1}\sum^{\infty}_{m_{i}=0}\alpha_{i}\big]\notag\\
&\times&\big[\bigotimes^{n}_{i=1}|m_{i}\rangle_{out}|m_{i}\rangle_{in}\big]+\cdots+(1-e^{-\frac{\omega}{T}})^{\frac{n+1}{2}}|0\rangle_{_{N}}|0\rangle_{_{N-1}}\cdots|0\rangle_{_{n+1}}\notag\\
&\times&\big[\prod^{n}_{i=2}\sum^{\infty}_{m_{i}=0}\alpha_{i}\big]\big[\bigotimes^{n}_{i=2}|m_{i}\rangle_{out}|m_{i}\rangle_{in}\big]\big[\beta_{m_1}|m+1_{1}\rangle_{out}|m_{1}\rangle_{in}\big]\bigg\}.
\end{eqnarray}

After a series of detailed calculations, the density operator of the N-qubit W state for the bosonic field is obtained as $\rho(W^{B}_{N})$ (see Appendix A for details). Using Eq.(\ref{S34}), we derive the analytical expression for the coherence $C^{B}(W)$ between accessible and inaccessible modes as
\begin{eqnarray}\label{S39}
C^{B}(W)&=&\frac{1-p}{N}\bigg\{x(x-1)\bigg[(1-e^{\frac{-\omega}{T}})^{\frac{3}{2}}\sum^{\infty}_{i=0}(e^{\frac{-n\omega}{2T}})^{2i}\sqrt{i+1}\bigg]^{2}\notag\\
&+&z(z-1)\bigg[(1-e^{\frac{-\omega}{T}})^{\frac{3}{2}}\sum^{\infty}_{i=0}(e^{\frac{-n\omega}{2T}})^{2i+1}\sqrt{i+1}\bigg]^{2}\notag\\
&+&2x(N-x-z)\bigg[(1-e^{\frac{-\omega}{T}})^{\frac{3}{2}}\sum^{\infty}_{i=0}(e^{\frac{-n\omega}{2T}})^{2i}\sqrt{i+1}\bigg]\notag\\
&+&2z(N-x-z)\bigg[(1-e^{\frac{-\omega}{T}})^{\frac{3}{2}}\sum^{\infty}_{i=0}(e^{\frac{-n\omega}{2T}})^{2i+1}\sqrt{i+1}\bigg]\notag\\
&+&(N-x-z)(N-x-z-1)\bigg\}.
\end{eqnarray}
The analytical expressions derived in this work involve finite or rapidly convergent series, and their numerical evaluation scales polynomially with the number of qubits. For large $N$, the symmetry of GHZ and W states allows analytical truncation, ensuring good scalability. Therefore, the present framework remains computationally tractable even for $N\ge100$.

\begin{figure}
\begin{minipage}[t]{0.5\linewidth}
\centering
\includegraphics[width=3.0in,height=5.2cm]{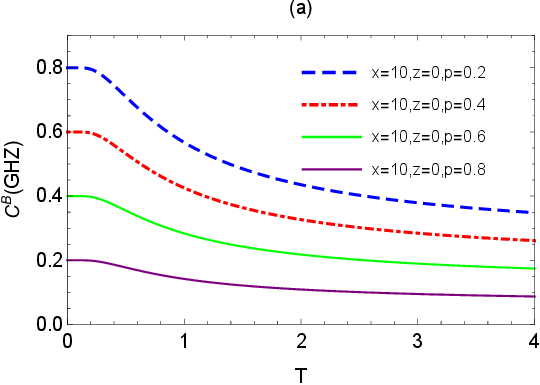}
\label{fig1a}
\end{minipage}%
\begin{minipage}[t]{0.5\linewidth}
\centering
\includegraphics[width=3.0in,height=5.2cm]{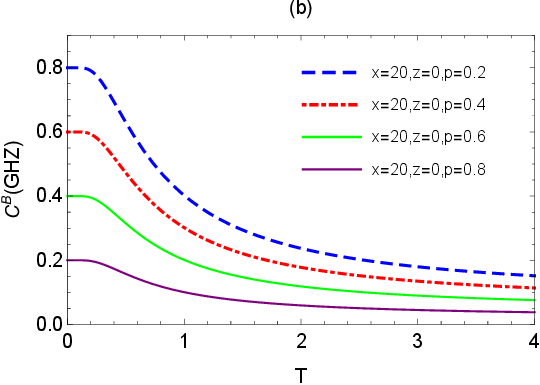}
\label{fig1b}
\end{minipage}%

\begin{minipage}[t]{0.5\linewidth}
\centering
\includegraphics[width=3.0in,height=5.2cm]{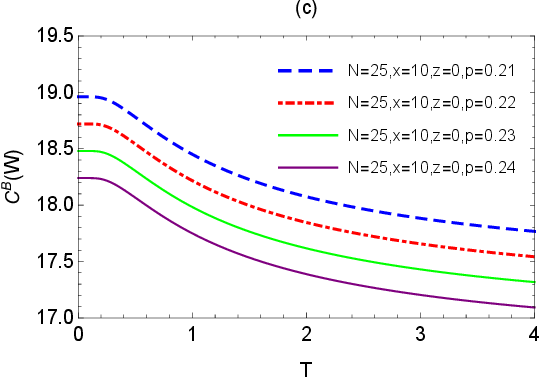}
\label{fig1c}
\end{minipage}%
\begin{minipage}[t]{0.5\linewidth}
\centering
\includegraphics[width=3.0in,height=5.2cm]{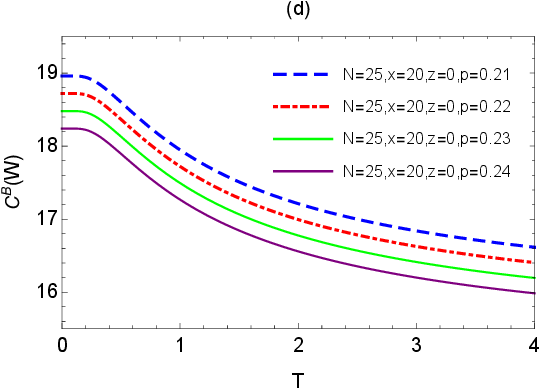}
\label{fig1d}
\end{minipage}%

\caption{The physically accessible coherence $C^{B}(GHZ)$ and $C^{B}(W)$ of bosonic field as functions of the Hawking temperature $T$ for different $p$ and $x$, where we have fixed $M=\omega=1$.}
\label{Fig1}
\end{figure}

In Fig.\ref{Fig1}, we illustrate how the multiqubit physically accessible coherence, $C^{B}(GHZ)$ and $C^{B}(W)$, changes with the Hawking temperature $T$ for different values of the mixing parameter $p$ and the number of accessible modes $x$. In Fig.\ref{Fig1}, we find that the physically accessible coherence of both mixed GHZ and W states decreases monotonically with increasing $T$, indicating that the thermal noise induced by the Hawking effect acts as a source of decoherence for the accessible part of the quantum system. Furthermore, in Fig.\ref{Fig1}, we also find that the coherence of both mixed GHZ and W states decrease as the mixing parameter $p$ increases, consistent with its role in quantifying the degree of depolarizing noise. A comparison between the mixed GHZ and W states across all configurations reveals that the mixed W state consistently retains more coherence than the mixed GHZ state in curved spacetime. Moreover, increasing the number of accessible bosonic modes $x$ near the event horizon leads to a further reduction in coherence, suggesting that proximity to the Hawking effect of the black hole enhances the decoherence effect on the physically accessible subsystem.

\begin{figure}
\begin{minipage}[t]{0.5\linewidth}
\centering
\includegraphics[width=3.0in,height=5.2cm]{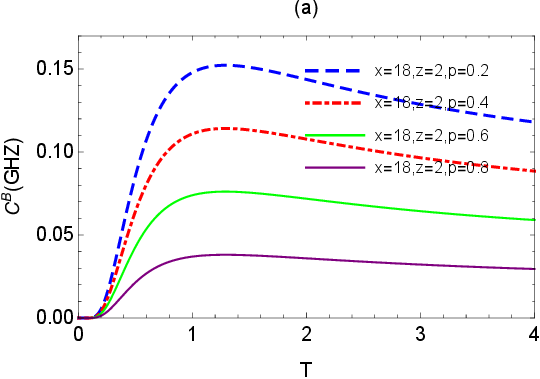}
\label{fig1e}
\end{minipage}%
\begin{minipage}[t]{0.5\linewidth}
\centering
\includegraphics[width=3.0in,height=5.2cm]{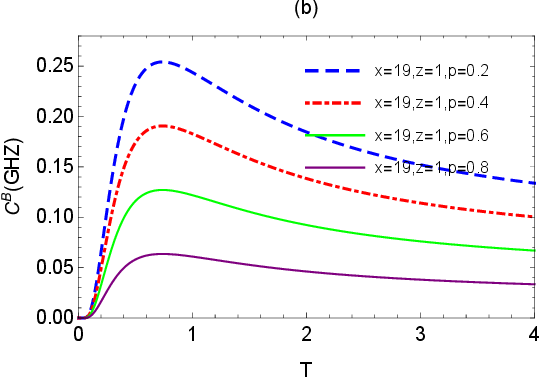}
\label{fig1f}
\end{minipage}%

\begin{minipage}[t]{0.5\linewidth}
\centering
\includegraphics[width=3.0in,height=5.2cm]{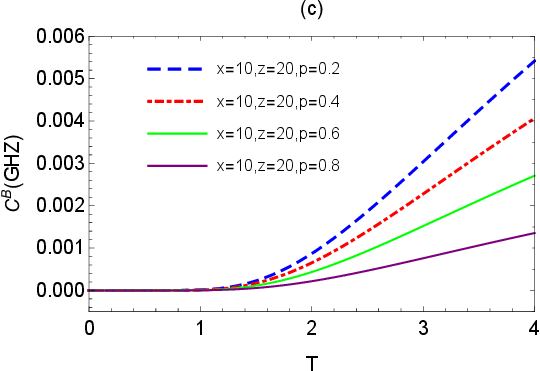}
\label{fig1i}
\end{minipage}%
\begin{minipage}[t]{0.5\linewidth}
\centering
\includegraphics[width=3.0in,height=5.2cm]{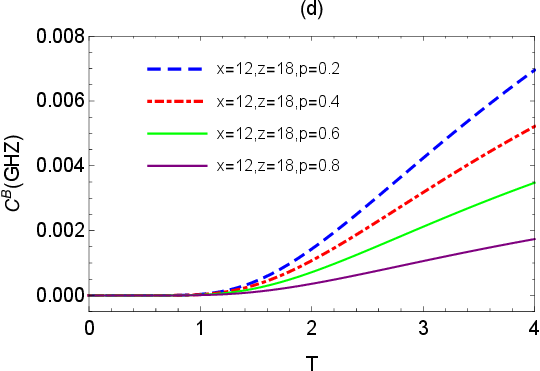}
\label{fig1j}
\end{minipage}%

\begin{minipage}[t]{0.5\linewidth}
\centering
\includegraphics[width=3.0in,height=5.2cm]{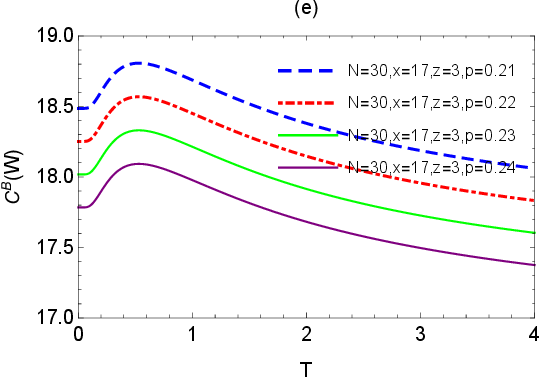}
\label{fig1g}
\end{minipage}%
\begin{minipage}[t]{0.5\linewidth}
\centering
\includegraphics[width=3.0in,height=5.2cm]{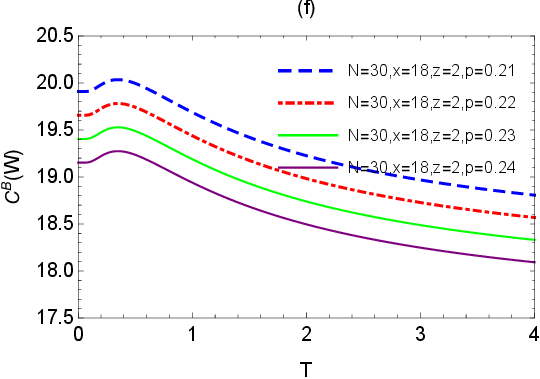}
\label{fig1h}
\end{minipage}%

\begin{minipage}[t]{0.5\linewidth}
\centering
\includegraphics[width=3.0in,height=5.2cm]{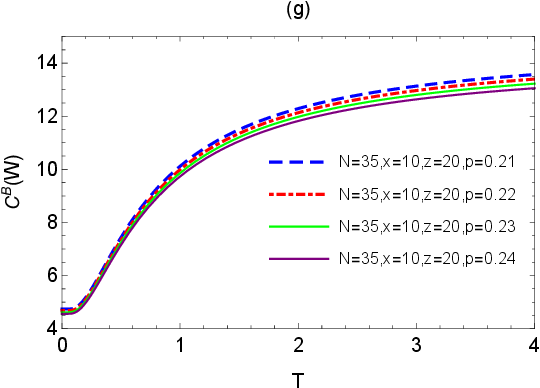}
\label{fig1k}
\end{minipage}%
\begin{minipage}[t]{0.5\linewidth}
\centering
\includegraphics[width=3.0in,height=5.2cm]{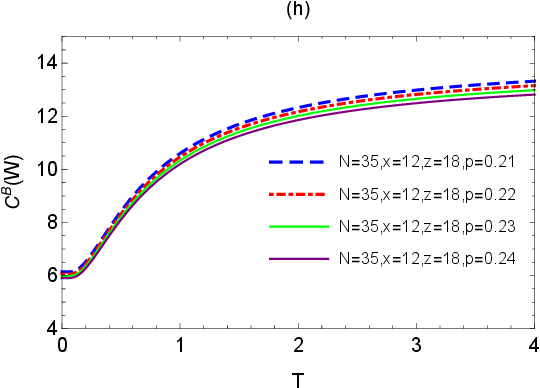}
\label{fig1l}
\end{minipage}%
\caption{The physically inaccessible coherence $C^{B}(GHZ)$ and $C^{B}(W)$ of bosonic field as functions of the Hawking temperature $T$ for different $N$, $p$, $x$, and $n$, with fixed parameters $M=\omega=1$.}
\label{Fig2}
\end{figure}

In Fig.\ref{Fig2},  we present the behavior of the physically inaccessible coherence, $C^{B}(GHZ)$ and $C^{B}(W)$,  as functions of the Hawking temperature $T$ for different values of $N$, $p$, $x$, and $n$. From Fig.\ref{Fig2}, it can be observed that for different values of $p$, as the Hawking temperature $T$ increases, the physically inaccessible coherence exhibits two different behaviors: monotonically increasing or non-monotonically increasing. This suggests that the behavior of inaccessible coherence is highly sensitive to the balance between the number of accessible modes $x$ and inaccessible modes $z$. Furthermore, as the mixing parameter $p$ increases, the coherence of both GHZ and W states consistently decreases. Since $p$ characterizes the degree of noise or decoherence in the system, higher values of  $p$ diminish the off-diagonal elements of the density matrix, which directly reduces quantum coherence and leads to a degradation of quantum properties.

\begin{figure}
\begin{minipage}[t]{0.5\linewidth}
\centering
\includegraphics[width=3.0in,height=5.2cm]{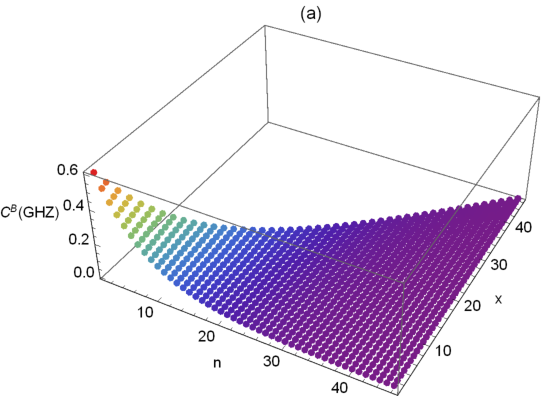}
\label{fig2a}
\end{minipage}%
\begin{minipage}[t]{0.5\linewidth}
\centering
\includegraphics[width=3.0in,height=5.2cm]{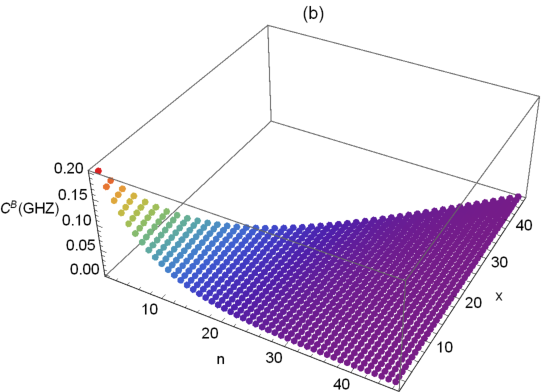}
\label{fig2b}
\end{minipage}%

\begin{minipage}[t]{0.5\linewidth}
\centering
\includegraphics[width=3.0in,height=5.2cm]{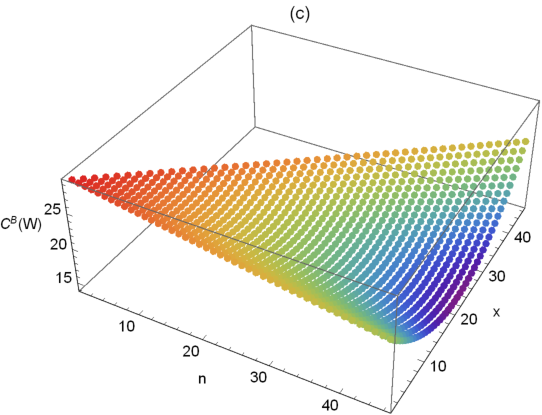}
\label{fig2c}
\end{minipage}%
\begin{minipage}[t]{0.5\linewidth}
\centering
\includegraphics[width=3.0in,height=5.2cm]{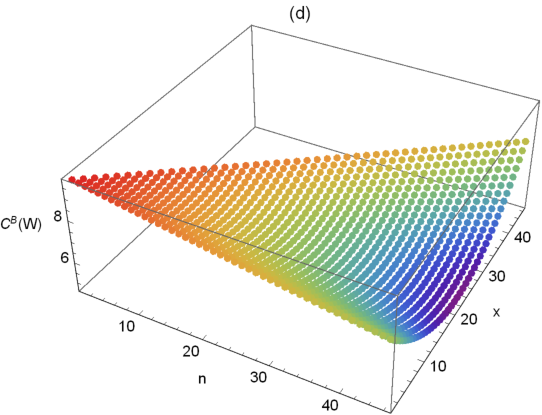}
\label{fig2d}
\end{minipage}%

\caption{Quantum coherence $C^{B}(GHZ)$ and $C^{B}(W)$ of bosonic field as functions of $x$ and $n$, where we have fixed $M=\omega=1$, and $T=20$. In Fig (a) and (c), the mixing parameter is set to $p=0.2$, while in Fig (b) and (d), it is fixed at $p=0.5$.}
\label{Fig3}
\end{figure}

In Fig.\ref{Fig3}, we plot the multiqubit coherence of bosonic field  as functions of the number of modes $n$ near event horizon and physically accessible modes $x$ for different $p$ in the context of the Schwarzschild black hole. From Fig.\ref{Fig3}, we can observe that when $x=n$ (the diagonal), it represents physically accessible coherence, where the coherence of the GHZ and W states monotonically decreases with the increase of $x$. In Fig.\ref{Fig3} (a) and (b), we can see that the coherence of the GHZ state monotonically decreases with increasing $n$ for a fixed $x$.
 A comparison between Fig.\ref{Fig3} (a) and (b) further reveals that as $p$ increases, the coherence of the GHZ state diminishes. This demonstrates that depolarization noise is a key mechanism responsible for quantum coherence degradation in the black hole. Meanwhile, Fig.\ref{Fig3} (c) and (d) show that, for a fixed $n$, the coherence of the W state first decreases to a minimum and then increases to a fixed value as $x$
increases. Additionally, a comparison between Fig.\ref{Fig3} (c) and (d) indicates that the coherence of the W state decreases monotonically with increasing $p$. Notably, for a fixed $n$, the coherence of the GHZ state remains constant as $x$ increases, whereas the W-state coherence exhibits non-monotonic behavior.
 When $T\rightarrow\infty$, we can obtain
\begin{equation}\nonumber
\begin{aligned}
\lim_{T\rightarrow\infty}\bigg[\big(1-e^{\frac{-\omega}{T}}\big)^{\frac{3}{2}}\sum^{\infty}_{i=0}\sqrt{i+1}\big(e^{\frac{-m\omega}{2T}}\big)^{2i}\bigg]\notag\\
=\lim_{T\rightarrow\infty}\bigg[\big(1-e^{\frac{-\omega}{T}}\big)^{\frac{3}{2}}\sum^{\infty}_{i=0}\sqrt{i+1}\big(e^{\frac{-m\omega}{2T}}\big)^{2i+1}\bigg]=\frac{\sqrt{\pi}}{2}.
\end{aligned}
\end{equation}
Therefore, from Eq.(\ref{S35}), the coherence of the mixed GHZ state for bosonic field is
\begin{equation}\nonumber
\begin{aligned}
C^{B}(GHZ)=\big(1-p\big)\lim_{T\rightarrow\infty}\bigg[\big(1-e^{\frac{-\omega}{T}}\big)^{\frac{3}{2}}\sum^{\infty}_{i=0}\sqrt{i+1}\big(e^{\frac{-m\omega}{2T}}\big)^{2i}\bigg]^{n}=\big(1-p\big)\bigg(\frac{\sqrt{\pi}}{2}\bigg)^{n}.
\end{aligned}
\end{equation}
This shows that the calculation results are consistent with the response of Fig.\ref{Fig3} (a) and (b).

\subsection{Multiqubit coherence of  mixed GHZ and W states for fermionic field}
Similar to the GHZ state of bosonic field, we employ the Schwarzschild modes defined by Eqs.(\ref{S24}) and (\ref{S25}) to reformulate Eq.(\ref{S27}) for fermionic field
\begin{eqnarray}\label{S40}
GHZ^{F}_{123\ldots N}&&=\frac{1}{\sqrt{2}}\bigg\{\bigotimes^{n}_{i=1}\overbrace{\big(|0\rangle_{N}|0\rangle_{N-1}\cdots|0\rangle_{n+1}\big)}^{|\bar{0}\rangle}\big[(e^{-\frac{\omega}{T}}+1)^{-\frac{1}{2}}|0_{i}\rangle_{out}|0_{i}\rangle_{in}\notag\\
&&+(e^{\frac{\omega}{T}}+1)^{-\frac{1}{2}}|1_{i}\rangle_{out}|1_{i}\rangle_{in}\big]+\bigotimes^{n}_{i=1}\overbrace{\big(|1\rangle_{N}|1\rangle_{N-1}\cdots|1\rangle_{n+1}\big)}^{|\bar{1}\rangle}\notag\\
&&\big[|1_{i}\rangle_{out}|0_{i}\rangle_{in}\big]\bigg\}.
\end{eqnarray}
By introducing noise and applying  Eq.(\ref{S29}), we derive the density operator of the  mixed
N-qubit GHZ state as
\begin{eqnarray}\label{S41}
\mathcal{\rho}(GHZ_{N}^{F})=\mathcal{\rho}^{GHZ_{N}^{F}}_{diag}+\mathcal{\rho}^{GHZ_{N}^{F}}_{non-diag},
\end{eqnarray}
where the non-diagonal part of the density operator for fermionic field is given by
\begin{eqnarray}\label{S42}
\mathcal{\rho}^{GHZ_{N}^{F}}_{non-diag}&=&\frac{1-p}{2}(\frac{1}{\sqrt{e^{\frac{-\omega}{T}}+1}})^{x}(\frac{1}{\sqrt{e^{\frac{\omega}{T}}+1}})^{z}\bigg\{|\overline{0}\rangle\langle\overline{1}|[\bigotimes^{x}_{i=1}(|0_{i}\rangle_{out}\langle1_{i}|)]\notag\\
&\times&[\bigotimes^{z}_{j=1}(|1_{j}\rangle_{in}\langle0_{j}|)]+|\overline{1}\rangle\langle\overline{0}|[\bigotimes^{x}_{i=1}(|1_{i}\rangle_{out}\langle0_{i}|)]\times[\bigotimes^{z}_{j=1}(|0_{j}\rangle_{in}\langle1_{j}|)]\bigg\}.
\end{eqnarray}

Through direct calculation, we obtain the analytical expression for the coherence of the mixed GHZ state in fermionic field as
\begin{eqnarray}\label{S43}
C^{F}(GHZ)=(1-p)(e^{\frac{-\omega}{T}}+1)^{-\frac{x}{2}}(e^{\frac{\omega}{T}}+1)^{-\frac{z}{2}}.
\end{eqnarray}

Based on the Eqs.(\ref{S24}), (\ref{S25}) and (\ref{S28}), the wave function of the W state for fermionic field can be rewritten as
\begin{eqnarray}\label{S44}
W^{F}_{123\ldots N+n}
&=&\frac{1}{\sqrt{N}}\bigg\{\bigotimes^{n}_{i=1}|1\rangle_{_{N}}|0\rangle_{_{N-1}}\cdots|0\rangle_{n
+1}\big[(e^{-\frac{\omega}{T}}+1)^{-\frac{1}{2}}|0_{i}\rangle_{out}|0_{i}\rangle_{in}\notag\\
&+&(e^{\frac{\omega}{T}}+1)^{-\frac{1}{2}}|1_{i}\rangle_{out}|1_{i}\rangle_{in}\big]+\bigotimes^{n}_{i=1}|0\rangle_{_{N}}|1\rangle_{_{N-1}}\cdots|0\rangle_{n+1}\notag\\
&\times&\big[(e^{-\frac{\omega}{T}}+1)^{-\frac{1}{2}}|0_{i}\rangle_{out}|0_{i}\rangle_{in}+(e^{\frac{\omega}{T}}+1)^{-\frac{1}{2}}|1_{i}\rangle_{out}|1_{i}\rangle_{in}\big]\notag\\
&+&\cdots+\bigotimes^{n}_{i=2}|0\rangle_{N}|0\rangle_{N-1}\cdots|0\rangle_{n+1}\big[(e^{-\frac{\omega}{T}}+1)^{-\frac{1}{2}}|0_{i}\rangle_{out}|0_{i}\rangle_{in}\notag\\
&+&(e^{\frac{\omega}{T}}+1)^{-\frac{1}{2}}|1_{i}\rangle_{out}|1_{i}\rangle_{in}\big]\big[|1_{1}\rangle_{out}|1_{1}\rangle_{in}\big]\bigg\}.
\end{eqnarray}
Similarly, the density operator for the N-qubit W state of the fermionic field is derived as $\rho(W^{F}_{N})$ (for details please see Appendix B).  Through tedious and direct calculations, the coherence of the mixed W state between the physically accessible and inaccessible modes can be calculated as
\begin{eqnarray}\label{S37}
C^{F}(W)&=&\frac{1-p}{N}\bigg[x(x-1)\big(e^{\frac{-\omega}{T}}+1\big)^{-1}+z(z-1)\big(e^{\frac{\omega}{T}}+1\big)^{-1}\notag\\
&+&2p(N-x-z)\big(e^{\frac{-\omega}{T}}+1\big)^{-\frac{1}{2}}+2q(N-x-z)\big(e^{\frac{\omega}{T}}+1\big)^{-\frac{1}{2}}\notag\\
&+&(N-x-z)(N-x-z-1)\bigg].
\end{eqnarray}

\begin{figure}
\begin{minipage}[t]{0.5\linewidth}
\centering
\includegraphics[width=3.0in,height=5.2cm]{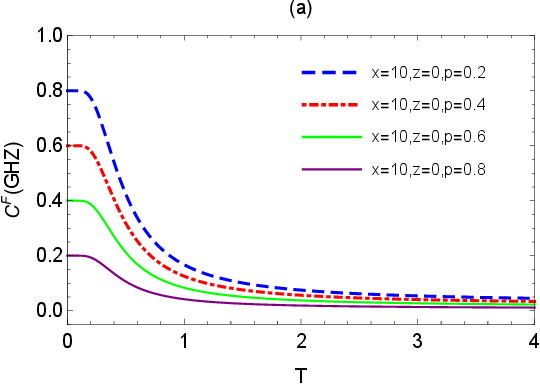}
\label{fig3a}
\end{minipage}%
\begin{minipage}[t]{0.5\linewidth}
\centering
\includegraphics[width=3.0in,height=5.2cm]{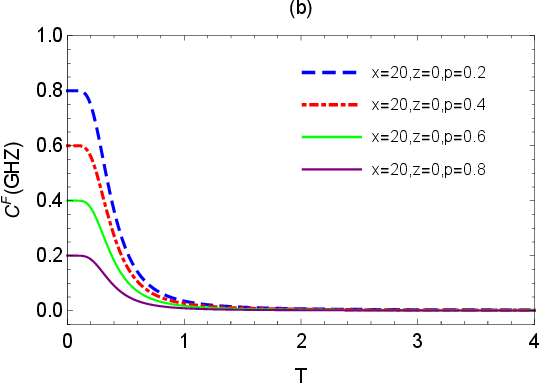}
\label{fig3b}
\end{minipage}%

\begin{minipage}[t]{0.5\linewidth}
\centering
\includegraphics[width=3.0in,height=5.2cm]{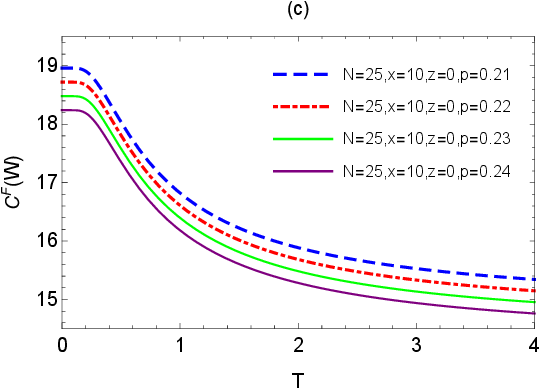}
\label{fig3c}
\end{minipage}%
\begin{minipage}[t]{0.5\linewidth}
\centering
\includegraphics[width=3.0in,height=5.2cm]{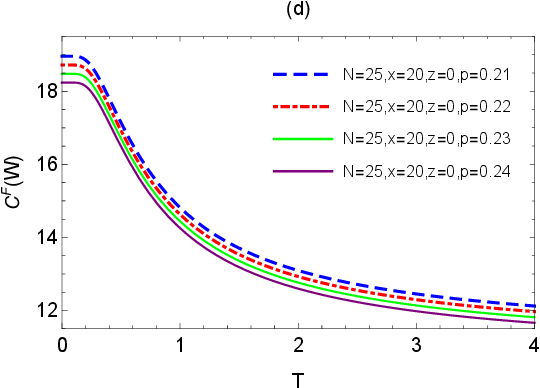}
\label{fig3d}
\end{minipage}%

\caption{The physically accessible quantum coherence measures $C^{F}(GHZ)$ and $C^{F}(W)$ of fermionic field are shown as functions of the Hawking temperature $T$, for varying mixing parameter $p$ and  parameter $x$, while keeping $M=\omega=1$ fixed.}
\label{Fig4}
\end{figure}

Fig.\ref{Fig4} presents the physically accessible coherence of the mixed GHZ and W states for fermionic field as  functions of Hawking temperature $T$, from which we observe that:
(i) the physically accessible coherence of both GHZ and W states decreases monotonically with increasing Hawking temperature $T$;
(ii) the coherence shows a monotonic decrease with respect to the mixed parameter $p$;
(iii) the coherence decreases as the parameter $x$ increases;
(iv) under identical conditions, the physically accessible coherence of the mixed W state is consistently greater than that of the mixed GHZ state.

\begin{figure}
\begin{minipage}[t]{0.5\linewidth}
\centering
\includegraphics[width=3.0in,height=5.2cm]{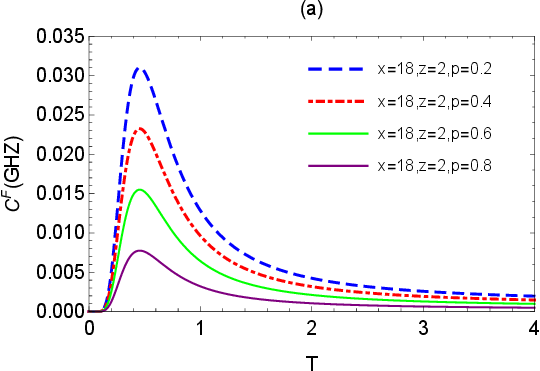}
\label{fig3e}
\end{minipage}%
\begin{minipage}[t]{0.5\linewidth}
\centering
\includegraphics[width=3.0in,height=5.2cm]{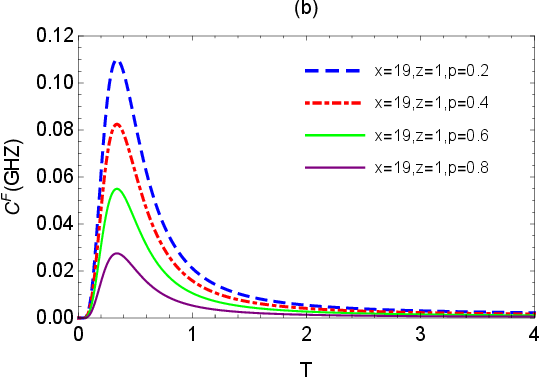}
\label{fig3f}
\end{minipage}%

\begin{minipage}[t]{0.5\linewidth}
\centering
\includegraphics[width=3.0in,height=5.2cm]{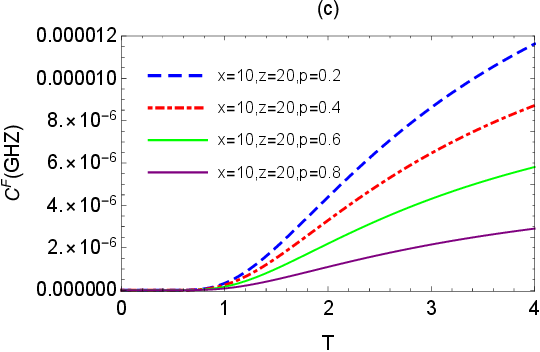}
\label{fig3i}
\end{minipage}%
\begin{minipage}[t]{0.5\linewidth}
\centering
\includegraphics[width=3.0in,height=5.2cm]{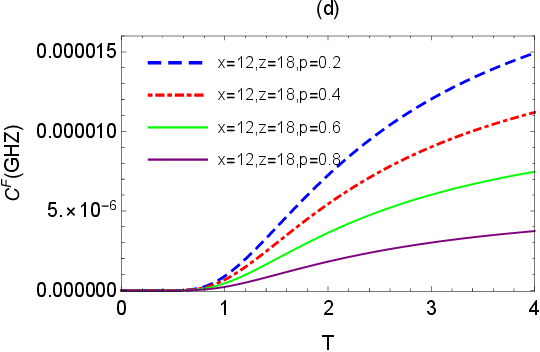}
\label{fig3j}
\end{minipage}%

\begin{minipage}[t]{0.5\linewidth}
\centering
\includegraphics[width=3.0in,height=5.2cm]{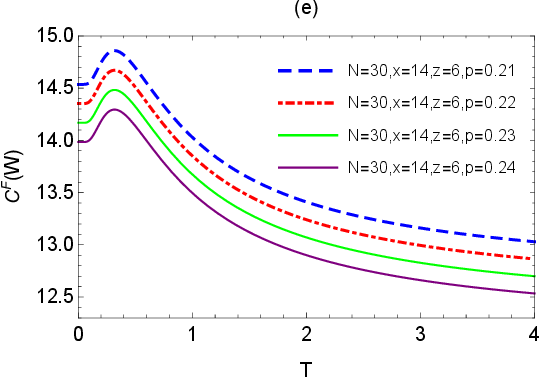}
\label{fig3g}
\end{minipage}%
\begin{minipage}[t]{0.5\linewidth}
\centering
\includegraphics[width=3.0in,height=5.2cm]{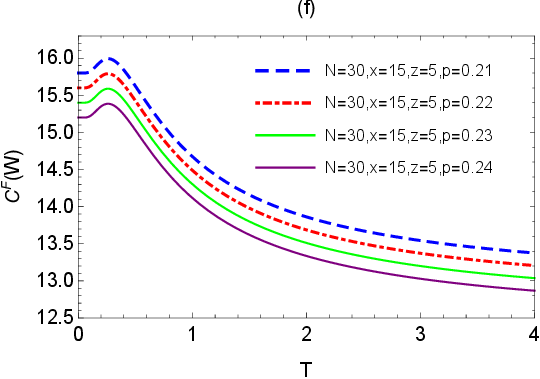}
\label{fig3h}
\end{minipage}%

\begin{minipage}[t]{0.5\linewidth}
\centering
\includegraphics[width=3.0in,height=5.2cm]{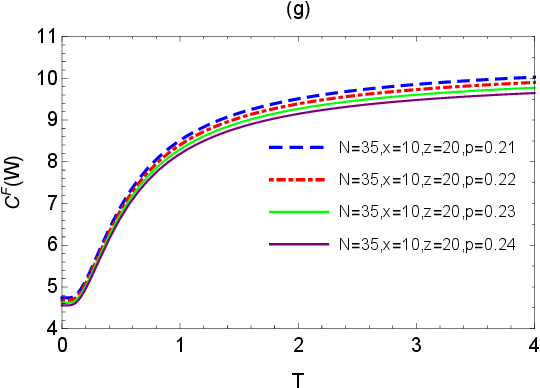}
\label{fig3k}
\end{minipage}%
\begin{minipage}[t]{0.5\linewidth}
\centering
\includegraphics[width=3.0in,height=5.2cm]{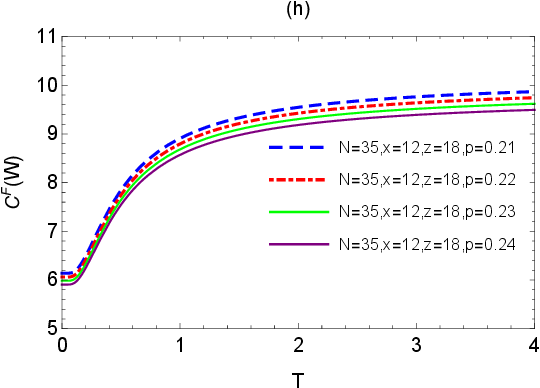}
\label{fig3l}
\end{minipage}%
\caption{The physically inaccessible quantum coherence measures $C^{F}(GHZ)$ and $C^{F}(W)$ of fermionic field are shown as functions of the Hawking temperature $T$, for varying $N$, $p$, $x$, and $n$, where we have fixed $M=\omega=1$.}
\label{Fig5}
\end{figure}

Fig.\ref{Fig5} displays the physically inaccessible coherence of the mixed GHZ and W states versus Hawking temperature $T$, revealing that:
(1) the coherence exhibits either monotonic or non-monotonic growth with increasing $T$, depending on the relative proportion of physically accessible and inaccessible modes;
(2) in all scenarios, the mixed W state consistently retains higher coherence than the mixed GHZ state;
(3) the coherence decreases with increasing inaccessible modes $z$;
(4) both states show diminishing coherence as the mixing parameter $p$ increases.

\begin{figure}
\begin{minipage}[t]{0.5\linewidth}
\centering
\includegraphics[width=3.0in,height=5.2cm]{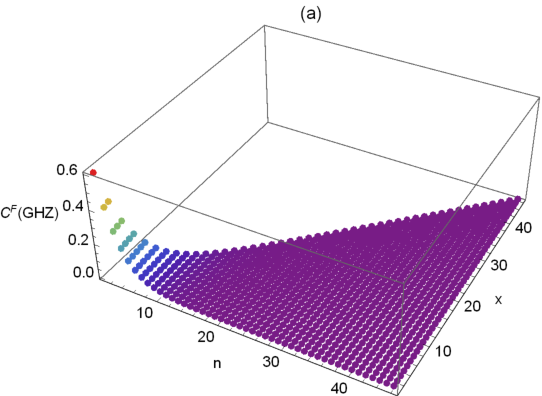}
\label{fig4a}
\end{minipage}%
\begin{minipage}[t]{0.5\linewidth}
\centering
\includegraphics[width=3.0in,height=5.2cm]{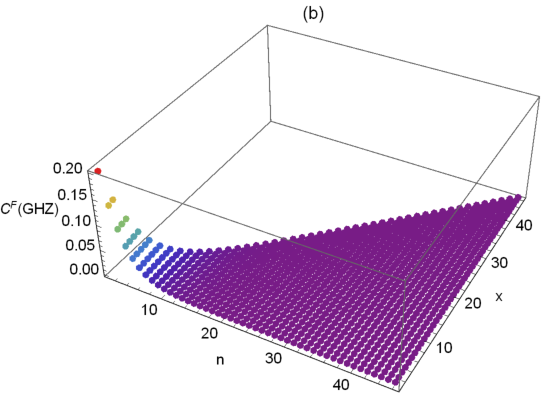}
\label{fig4b}
\end{minipage}%

\begin{minipage}[t]{0.5\linewidth}
\centering
\includegraphics[width=3.0in,height=5.2cm]{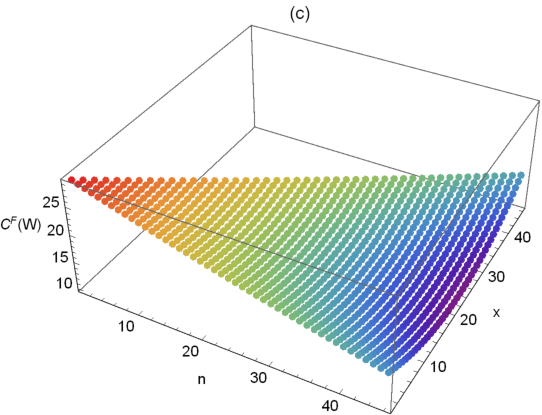}
\label{fig4c}
\end{minipage}%
\begin{minipage}[t]{0.5\linewidth}
\centering
\includegraphics[width=3.0in,height=5.2cm]{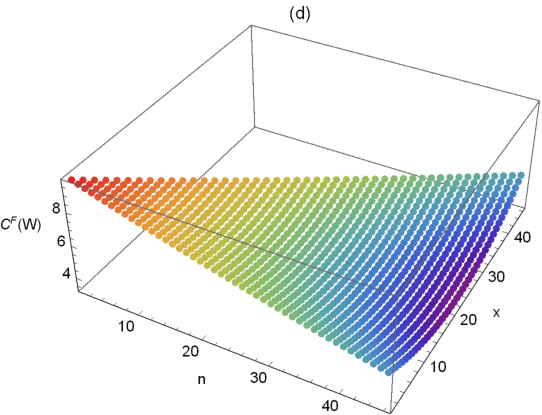}
\label{fig4d}
\end{minipage}%

\caption{Quantum coherence $C^{F}(GHZ)$ and $C^{F}(W)$ of fermionic field are analyzed as functions of $x$ and $n$, where we have fixed $M=\omega=1$, and $T=20$. The mixing parameter is set to $p=0.2$ in Figs (a) and (c), while in Figs (b) and (d), it is fixed at $p=0.5$.}
\label{Fig6}
\end{figure}

Fig.\ref{Fig6} presents the multiqubit coherence of both mixed GHZ and W states for fermionic field as functions of the number of modes $n$ near the event horizon and physically accessible modes $x$ for different values of $p$ in Schwarzschild spacetime. Analysis of the four subplots reveals three key findings: (i) when $x = n$, the coherence of both states decreases with increasing $x$;
(ii) for fixed $x$ and $n$, the coherence of both states monotonically decreases with increasing $p$; (iii) under the extreme condition $T\rightarrow\infty$ and fixed $n$, the coherence of the mixed GHZ state remains invariant with increasing $x$, whereas the coherence of the mixed W state first decreases to a minimum and then increases toward a stable value as $x$ increases.
From Eq.(\ref{S43}), it can be shown that in the high-temperature limit $T\rightarrow\infty$, the following identity holds
\begin{equation}\nonumber
\begin{aligned}
\lim_{ T\rightarrow\infty}(e^{\frac{-\omega}{T}}+1)^{-\frac{1}{2}}=\lim_{T\rightarrow\infty}(e^{\frac{\omega}{T}}+1)^{-\frac{1}{2}}=\frac{\sqrt{2}}{2}.
\end{aligned}
\end{equation}
This analytical result confirms that, in the limit $T\rightarrow\infty$, the coherence of the mixed GHZ state in fermionic field becomes independent of $x$. In this case, the coherence simplifies to
\begin{equation}
\lim_{T\rightarrow\infty}C^{F}(GHZ)=(1-p)\lim_{T\rightarrow\infty}(e^{\frac{-\omega}{T}}+1)^{-\frac{n}{2}}=(1-p)\bigg(\frac{\sqrt{2}}{2}\bigg)^{n}.
\end{equation}

\begin{figure}
\begin{minipage}[t]{0.5\linewidth}
\centering
\includegraphics[width=3.0in,height=5.2cm]{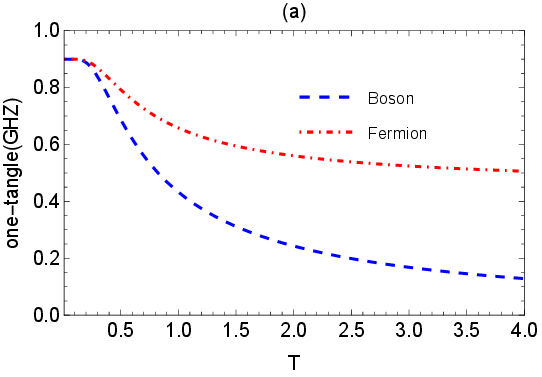}
\label{fig10a}
\end{minipage}%
\begin{minipage}[t]{0.5\linewidth}
\centering
\includegraphics[width=3.0in,height=5.2cm]{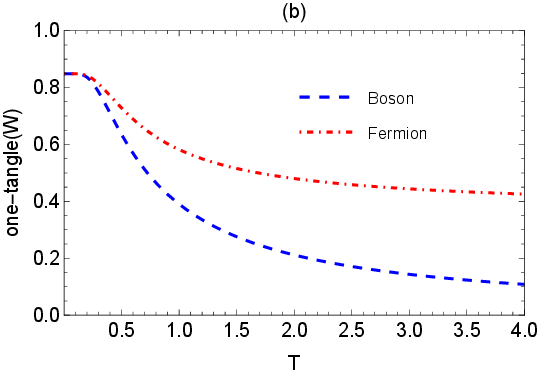}
\label{fig10b}
\end{minipage}%

\begin{minipage}[t]{0.5\linewidth}
\centering
\includegraphics[width=3.0in,height=5.2cm]{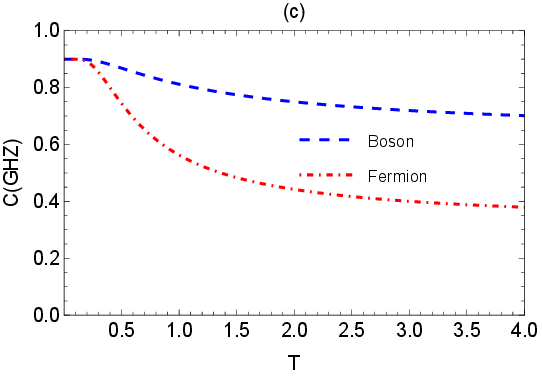}
\label{fig10c}
\end{minipage}%
\begin{minipage}[t]{0.5\linewidth}
\centering
\includegraphics[width=3.0in,height=5.2cm]{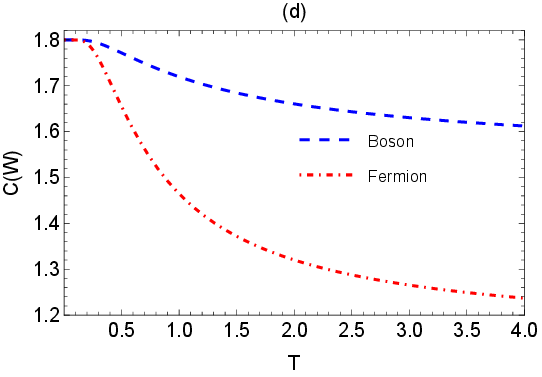}
\label{fig10d}
\end{minipage}%

\caption{The coherence and entanglement of bosonic and fermionic fields as functions of the Hawking temperature $T$ for fixed $M=\omega=1$ and $p=0.1$.}
\label{Fig10}
\end{figure}

We will conduct a systematic and visually comparative quantitative analysis of the quantum coherence and entanglement properties of GHZ and W states for bosonic and fermionic fields in Schwarzschild spacetime. Given the complexity of entanglement structures in bosonic fields, we adopt tripartite entanglement as the fundamental model for in-depth investigation. We plot the tripartite coherence and entanglement as functions of the Hawking temperature for mixed states in both bosonic and fermionic fields.
Fig.\ref{Fig10} shows how the Hawking temperature $T$  affects quantum entanglement and coherence of mixed GHZ and W states for bosonic and fermionic fields.  We can find that  coherence of bosonic field consistently exceeds coherence of fermionic field in Schwarzschild spacetime and remains robust even in the presence of an extreme black hole, whereas the fermionic field demonstrates more robust entanglement than the bosonic field. Regarding different initial quantum states, the entanglement of the GHZ state is generally stronger than that of the W state. However, the physically accessible multiqubit coherence of the W state is consistently greater than that of the GHZ state in curved spacetime. These findings suggest that the optimal choice of multiqubit quantum resources for relativistic quantum information tasks depends critically on both the type of field (bosonic or fermionic) and the nature of the initial quantum state.

In recent years, analog black hole systems and simulated curved spacetimes have gained significant attention as controllable laboratory platforms for investigating quantum gravitational effects. A variety of experimental realizations, such as optical analogs, acoustic black holes in flowing fluids, and Bose–Einstein condensate horizons, have successfully reproduced Hawking-like radiation and quantum correlations under curved-spacetime conditions \cite{qhx1,qhx2,qhx3,qhx4,qhx5,qhx6}.  These analog systems provide an important bridge between theoretical predictions and experimental observability, allowing key aspects of relativistic quantum information, such as decoherence, entanglement degradation, and coherence redistribution, to be tested under well-controlled settings. Therefore, our investigation of multiqubit coherence and its redistribution near an event horizon may provide theoretical guidance for interpreting and designing future analog-gravity experiments.

\section{Conclusions}
In this study, we have explored the multiqubit coherence of the mixed GHZ and W states for bosonic and fermionic fields near the event horizon of a Schwarzschild black hole, revealing key insights into their behavior under relativistic conditions. The coherence of both states depends critically on the initial mixing parameter $p$ and Hawking temperature $T$, with higher $p$ and $T$ leading to greater decoherence due to depolarizing noise and thermal effects.  Notably, the physically inaccessible coherence of both mixed states exhibits either monotonic or non-monotonic growth with  Hawking temperature, depending on the balance between accessible and inaccessible modes.
Despite its relatively weaker entanglement, the W state maintains quantum coherence more effectively than the GHZ state in the presence of Hawking radiation, indicating a higher robustness under gravitational influence \cite{M108,M109}. Interestingly, bosonic fields exhibit stronger coherence than fermionic fields in Schwarzschild spacetime, whereas fermions maintain greater entanglement–a striking contrast that underscores the influence of particle statistics on quantum resources \cite{R1,R2,WDD4,RM1,RM3,RM5,B31,R15}. Additionally, for the W state, increasing the number of qubits enhances coherence but diminishes entanglement, highlighting a fundamental trade-off in relativistic quantum systems.
By comparing different quantum states, particle statistics, and types of quantum resources, these insights provide how relativistic effects fundamentally reshape resource trade-offs, offering strategic insights for optimizing quantum information tasks in curved spacetime.

Although the present work is primarily theoretical, recent advances in experimental quantum technologies indicate that the physical scenarios considered here could be probed in realistic or analog settings. Quantum optical experiments using the Micius satellite have already tested quantum field behavior in curved spacetime by transmitting entangled photon pairs through different regions of Earth's gravitational potential, providing a direct experimental platform for investigating gravity induced decoherence and non-unitary effects \cite{qhx7}. Likewise,  atomic clock experiments have measured gravitational redshifts within millimeter-scale samples of ultracold strontium atoms, reaching sensitivities capable of probing quantum coherence at distinct gravitational potentials \cite{qhx8}. In parallel, the development of compact quantum microsatellites and satellite-based quantum key distribution networks demonstrates that quantum coherence and entanglement can be distributed across large gravitational gradients between ground and orbit \cite{qhx9}. These developments collectively suggest that analog implementations of relativistic quantum information processes, such as qubits subjected to varying gravitational fields or near an event horizon analog potentials, are becoming experimentally accessible. The theoretical framework presented in this work provides a foundation for interpreting these emerging experiments and for guiding the design of future analog gravity platforms capable of simulating qubit dynamics and coherence redistribution in curved spacetime.

\appendix
\onecolumngrid

\section{Density operator of W state of bosonic field }
We decompose the density operator of the system into diagonal and non-diagonal parts based on the
N-qubit mixed  W state defined in Eq.(\ref{S30}) within the asymptotically flat region
\begin{eqnarray}\label{S37}
\mathcal{\tilde{\rho}}^{B/F}(W)=\mathcal{\tilde{\rho}}^{W_{N}^{B/F}}_{diag}+\mathcal{\tilde{\rho}}^{W_{N}^{B/F}}_{non-diag},
\end{eqnarray}
where the non-diagonal density operator can be further decomposed into three components
\begin{eqnarray}\label{S38}
\mathcal{\tilde{\rho}}^{W_{N}^{B/F}}_{non-diag}=\mathcal{\tilde{\rho}}^{W_{N}^{B/F1}}_{non-diag}+\mathcal{\tilde{\rho}}^{W_{N}^{B/F2}}_{non-diag}+\mathcal{\tilde{\rho}}^{W_{N}^{B/F3}}_{non-diag}.
\end{eqnarray}
Here, we focus exclusively on the off-diagonal density operator, since in the Schwarzschild black hole background, all quantum coherence arises entirely from this off-diagonal component.
The off-diagonal density operator $\mathcal{\tilde{\rho}}^{W_{N}^{B1}}_{non-diag}$ of the mixed  W state for the bosonic field in the asymptotically flat region can be expressed as
\begin{eqnarray}\label{S412}
\mathcal{\tilde{\rho}}^{W_{N}^{B1}}_{non-diag}&=&\frac{1-p}{N}\big[\big(\sum_{j=n+1}^{N}\sum_{i=1}^{n}|0_{N}0_{N-1}\ldots0_{n+1}0_{n}0_{n-1}\ldots0_{i+1}1_{i}0_{i-1}\ldots0_{2}0_{1}\rangle
\notag\\
&&\bigotimes\langle0_{N}0_{N-1}\ldots0_{j+1}1_{j}0_{j-1}\ldots0_{n+1}0_{n}0_{n-1}\ldots0_{2}0_{1}|\big)+h.c.\big],
\end{eqnarray}
while the other two off-diagonal components read
\begin{eqnarray}\label{S413}
\mathcal{\tilde{\rho}}^{W_{N}^{B2}}_{non-diag}&=&\frac{1-p}{N}\big(\sum_{i,j=1(i\ne j)}^{n}|0_{N}0_{N-1}\ldots0_{n+1}0_{n}0_{n-1}\ldots0_{i+1}1_{i}0_{i-1}\ldots0_{2}0_{1}\rangle
\notag\\
&&\bigotimes\langle0_{N}0_{N-1}\ldots0_{n+1}0_{n}0_{n-1}\ldots0_{j+1}1_{j}0_{j-1}\ldots0_{2}0_{1}|\big),
\end{eqnarray}
\begin{eqnarray}\label{S414}
\mathcal{\tilde{\rho}}^{W_{N}^{B3}}_{non-diag}&=&\frac{1-p}{N}\big(\sum_{i,j=n+1(i\ne j)}^{N}|0_{N}0_{N-1}\ldots0_{i+1}1_{i}0_{i-1}\ldots0_{n+1}0_{n}0_{n-1}\ldots0_{2}0_{1}\rangle
\notag\\
&&\bigotimes\langle0_{N}0_{N-1}\ldots0_{j+1}1_{j}0_{j-1}\ldots0_{n+1}0_{n}0_{n-1}\ldots0_{2}0_{1}|\big).
\end{eqnarray}

Specifically, all particles with indices less than or equal to  $n$ are located near the event horizon of the black hole, while those with indices greater than  $n$ remain in the asymptotically flat region. Under the influence of Hawking radiation, the quantum states of particles hovering near the event horizon evolve from the Kruskal vacuum representing the global vacuum state into two-mode squeezed states that describe the entanglement between the interior and exterior modes of the horizon. Accordingly, the density operator of the  W state in Eq.(\ref{S412}) is rewritten as
\begin{eqnarray}\label{S39}
\mathcal{\rho}^{W_{N}^{B1}}_{non-diag}&=&\frac{1-p}{N}\bigg\{\overbrace{\big[\sum^{N-n}_{m=1}\bigotimes^{N}_{j=n+1,j\neq n+m}|0_{j}\rangle \langle 0_{j}||0_{n+m}\rangle \langle1_{n+m}|\big]}^{flat}\big[\big(1-e^{\frac{-\omega}{T}}\big)^{\frac{2n+1}{2}}\big]\notag\\
&\times&\big\{\big[\prod^{n}_{l=z+1}\sum^{\infty}_{n_{l}=0}\alpha_{n_{l}}^{2}\big]\big[\bigotimes^{n}_{l=z+1}|n_{l}\rangle_{out}\langle n_{l}|\big]\big[\prod^{z}_{i=1}\sum^{\infty}_{n_{i}=0}\alpha_{n_{i}}^{2}\big]\big[\bigotimes^{z}_{i=1(i\neq k)}|n_{i}\rangle_{in}\langle n_{i}|\big]\notag\\
&\times&\big[\sum^{\infty}_{n_{k}=0}\beta_{n_{k}}\gamma_{n_{k}}\big]\big[\bigotimes^{z}_{k=1}|n_{k}\rangle_{in}\langle n+1_{k}|\big]+\big[\prod^{n}_{l=z+1}\sum^{\infty}_{n_{l}=0}\alpha_{n_{l}}^{2}\big]\big[\bigotimes^{n}_{l=z+1,l\ne q}|n_{l}\rangle_{out}\langle n_{l}|\big]\notag\\
&\times&\big[\sum^{\infty}_{n_{q}=0}\alpha_{n_{q}}\beta_{n_{q}}\big]\big[\sum^{n}_{q=z+1}|n+1_{q}\rangle_{out}\langle n_{q}|\big]\big[\prod^{z}_{i=1}\sum^{\infty}_{n_{i}=0}\alpha_{n_{i}}^{2}\big]\big[\bigotimes^{z}_{i=1}|n_{i}\rangle_{in}\langle n_{i}|\big]\big\}+h.c.\bigg\},
\end{eqnarray}
while Eq.(\ref{S413}) and Eq.(\ref{S414}) can be rewritten as
\begin{eqnarray}\label{S40}
\mathcal{\rho}^{W_{N}^{B2}}_{non-diag}&=&\frac{1-p}{N}\bigg\{\overbrace{\big[\bigotimes^{N}_{j=n+1}|0_{j}\rangle\langle 0_{j}|\big]}^{flat}\big[\big(1-e^{\frac{-\omega}{T}}\big)^{n+1}\big]\big\{\bigg[\big[\prod^{n}_{l=z+1}\sum^{\infty}_{n_{l}=0}\alpha_{n_{l}}^{2}\big]\big[\bigotimes^{n}_{l=z+1}|n_{l}\rangle_{out}\langle n_{l}|\big]\notag\\
&\times&\big[\sum^{z}_{k=1}\sum^{\infty}_{n_{k}=0}\beta_{n_{k}}\gamma_{{n_{k}}}\big]\big[|n_{k}\rangle_{in}\langle n+1_{k}|\big]\big[\sum^{z}_{m=1}\sum^{\infty}_{n_{m}=0}\beta_{n_{m}}\gamma_{{n_{m}}}\big]\big[|n+1_{m}\rangle_{in}\langle n_{m}|\big]\notag\\
&\times&\big[\prod^{z}_{i=1}\sum^{\infty}_{n_{i}=0}\alpha_{n_{i}}^{2}\big]\big[\bigotimes^{z}_{i=1(i\neq k,m)} |n_{i}\rangle_{in}\langle n_{i}|\big]\bigg]+\bigg[\big[\sum^{n}_{p=z+1}\sum^{\infty}_{n_{p}=0}\alpha_{n_{p}}\beta_{n_{p}}\big]\big[|n+1_{p}\rangle_{out}\langle n_{p}|\big]\notag\\
&\times&\big[\prod^{n}_{l=z+1}\sum^{\infty}_{n_{l}=0(l\neq p)}\alpha_{n_{l}}^{2}\big]\big[\bigotimes^{n}_{l=z+1(l\neq p)}|n_{l}\rangle_{out}\langle n_{l}|\big]\big[\sum^{z}_{m=1}\sum^{\infty}_{n_{m}=0}\beta_{n_{m}}\gamma_{n_{m}}\big]\notag\\
&\times&\big[|n+1_{m}\rangle_{in}\langle n_{m}|\big]\big[\prod^{z}_{i=1}\sum^{\infty}_{n_{i}=0}\alpha_{n_{i}}^{2}\big]\big[\bigotimes^{z}_{i=1(i\neq m)}|n_{i}\rangle_{in}\langle n_{i}|\big]+h.c.\bigg]\notag\\
&+&\bigg[\big[\sum^{n}_{p=z+1}\sum^{\infty}_{n_{p}=0}\alpha_{n_{p}}\beta_{n_{p}}\big]\big[|n+1_{p}\rangle_{out}\langle n_{p}|\big]\big[\sum^{n}_{s=z+1,s\ne p}\sum^{\infty}_{n_{s}=0}\alpha_{n_{s}}\beta_{n_{s}}\big]\notag\\
&\times&\big[|n_{s}\rangle_{out}\langle n+1_{s}|\big]\big[\prod^{n}_{l=z+1}\sum^{\infty}_{n_{l}=0}\alpha_{n_{l}}^{2}\big]\big[\bigotimes^{n}_{l=z+1,l\ne s,p}|n_{l}\rangle_{out}\langle n_{l}|\big]\notag\\
&\times&\big[\prod^{z}_{i=1}\sum^{\infty}_{n_{i}=0}\alpha_{n_{i}}^{2}\big]\big[\bigotimes^{z}_{i=1}|n_{i}\rangle_{in}\langle n_{i}|\big]\bigg]\big\}\bigg\},
\end{eqnarray}
\begin{eqnarray}\label{S41}
\mathcal{\rho}^{W_{N}^{B3}}_{non-diag}&=&\frac{1-p}{N}\bigg\{\overbrace{\sum^{N-n}_{u,m=1,{u\ne m}}|1_{n+u}\rangle\langle 0_{n+u}||0_{n+m}\rangle\langle 1_{n+m}|\bigotimes^{N-n}_{j=n+1,j\neq n+u,n+m}|0_{j}\rangle\langle 0_{j}|}^{flat}\big[\big(1-e^{\frac{-\omega}{T}}\big)^{n}\big]\notag\\
&\times&\big\{\big[\prod^{n}_{l=z+1}\sum^{\infty}_{n_{l}=0}\alpha_{n_{l}}^{2}\big]\big[\bigotimes^{n}_{l=z+1}|n_{l}\rangle_{out}\langle n_{l}|\big]\big[\prod^{z}_{i=1}\sum^{\infty}_{n_{i}=0}\alpha_{n_{i}}^{2}\big]\big[\bigotimes^{z}_{i=1}|n_{i}\rangle_{in}\langle n_{i}|\big]\big\}\bigg\}.
\end{eqnarray}
To clarify the physical meaning in the above expressions, the label $\overbrace{flat}$ denotes particles residing in the asymptotically flat spacetime region, whereas the unlabeled terms correspond to the subsystem of $n$ particles near the event horizon of which $x$ are located in the physically accessible region and $z$ in the physically inaccessible region.

\section{Density operator of W state of fermionic field }
Under the influence of Hawking radiation, the off-diagonal density operator  $\mathcal{\tilde{\rho}}^{W_{N}^{F1}}_{non-diag}$ of fermionic field can be rewritten as
\begin{eqnarray}\label{S44}
\mathcal{\rho}^{W_{N}^{F1}}_{non-diag}&=&\frac{1-p}{N}\bigg\{\overbrace{\big[\sum^{N-n}_{m=1}\bigotimes^{N}_{j=n+1,j\neq n+m}|0_{j}\rangle \langle 0_{j}||0_{n+m}\rangle \langle1_{n+m}|\big]}^{flat}\big\{\big[\bigotimes^{n}_{l=z+1}\big(\tilde{a}^{2}|0_{l}\rangle_{out}\langle 0_{l}|\notag\\
&+&\tilde{b}^{2}|1_{l}\rangle_{out}\langle 1_{l}|\big)\big]\big[\sum^{z}_{k=1}\bigotimes^{z}_{i=1(i\neq k)}\big(\tilde{a}^{2}|0_{i}\rangle_{in}\langle 0_{i}|+\tilde{b}^{2}|1_{i}\rangle_{in}\langle 1_{i}|\big)\big(\tilde{b}|0_{k}\rangle_{in}\langle 1_{k}|\big)\big]\notag\\
&+&\big[\bigotimes^{n}_{l=z+1(l\neq q)}\big(\tilde{a}^{2}|0_{l}\rangle_{out}\langle 0_{l}|+\tilde{b}^{2}|1_{l}\rangle_{out}\langle 1_{l}|\big)\sum^{n}_{q=z+1}\big(\tilde{a}|1_{q}\rangle_{out}\langle 0_{q}|\big)\big]\notag\\
&\times&\big[\bigotimes^{z}_{i=1}\big(\tilde{a}^{2}|0_{i}\rangle_{in}\langle 0_{i}|+\tilde{b}^{2}|1_{i}\rangle_{in}\langle 1_{i}|\big)\big]\big\}+h.c.\bigg\},
\end{eqnarray}
where
\begin{eqnarray*}\nonumber
&&\tilde{a}=\frac{1}{\sqrt{e^{-\omega/T}+1}},\notag\\
&&\tilde{b}=\frac{1}{\sqrt{e^{\omega/T}+1}}.
\end{eqnarray*}
Similarly, the second off-diagonal component $\mathcal{\tilde{\rho}}^{W_{N}^{F2}}_{non-diag}$ can be represented as
\begin{eqnarray}\label{S45}
\mathcal{\rho}^{W_{N}^{F2}}_{non-diag}&=&\frac{1-p}{N}\bigg\{\overbrace{\big[\bigotimes^{N}_{j=n+1}|0_{j}\rangle \langle 0_{j}|\big]}^{flat}\big\{\bigg[\big[\bigotimes^{n}_{l=z+1}\big(\tilde{a}^{2}|0_{l}\rangle_{out}\langle 0_{l}|+\tilde{b}^{2}|1_{l}\rangle_{out}\langle 1_{l}|\big)\big]\notag\\
&\times&\big[\sum^{z}_{k=1(k\ne m)}\big(\tilde{b}|0_{k}\rangle_{in}\langle 1_{k}|\big)\big]\big[\sum^{z}_{m=1}\big(\tilde{b}|1_{m}\rangle_{in}\langle 0_{m}|\big)\big]\big[\bigotimes^{z}_{i=1(i\neq k,m)}\big(\tilde{a}^{2}|0_{i}\rangle_{in}\langle 0_{i}|+\tilde{b}^{2}|1_{i}\rangle_{in}\langle 1_{i}|\big)\big]\bigg]\notag\\
&+&\bigg[\big[\sum^{n}_{p=z+1}\big(\tilde{a}|1_{p}\rangle_{out}\langle 0_{p}|\big)\big]\big[\bigotimes^{n}_{l=z+1(l\ne p)}\big(\tilde{a}^{2}|0_{l}\rangle_{out}\langle 0_{l}|+\tilde{b}^{2}|1_{l}\rangle_{out}\langle 1_{l}|\big)\big]\big[\sum_{m=1}^{z}\big(\tilde{b}|1_{m}\rangle_{in}\langle 0_{m}|\big)\big]\notag\\
&\times&\big[\bigotimes^{z}_{i=1,i\ne m}\big(\tilde{a}^{2}|0_{i}\rangle_{in}\langle 0_{i}|+\tilde{b}^{2}|1_{i}\rangle_{in}\langle 1_{k}|\big)\big]+h.c.\bigg]+\bigg[\big[\sum^{n}_{p=z+1}\big(\tilde{a}|1_{p}\rangle_{out}\langle 0_{p}|\big)\big]\notag\\
&\times&\big[\sum^{n}_{s=z+1,s\ne p}\big(\tilde{a}|0_{s}\rangle_{out}\langle 1_{s}|\big)\big]\big[\bigotimes^{n}_{l=z+1(l\neq s,p)}\big(\tilde{a}^{2}|0_{l}\rangle_{out}\langle 0_{l}|+\tilde{b}^{2}|1_{l}\rangle_{out}\langle 1_{l}|\big)\big]\notag\\
&\times&\big[\bigotimes^{z}_{i=1}\big(\tilde{a}^{2}|0_{i}\rangle_{in}\langle 0_{i}|+\tilde{b}^{2}|1_{i}\rangle_{in}\langle 1_{k}|\big)\big]\bigg]\big\}\bigg\}.
\end{eqnarray}
Likewise, the third off-diagonal component $\mathcal{\tilde{\rho}}^{W_{N}^{F3}}_{non-diag}$ takes the form
\begin{eqnarray}\label{S46}
\mathcal{\rho}^{W_{N}^{F3}}_{non-diag}&=&\frac{1-p}{N}\bigg\{\overbrace{\big[\sum^{N-n}_{u,m=1(k\ne m)}|1_{n+u}\rangle\langle 0_{n+u}||0_{n+m}\rangle\langle 1_{n+m}|\sum^{N}_{j=n+1(j\neq n+m,n+u)}|0_{j}\rangle\langle0_{j}|\big]}^{flat}\notag\\
&\times&\big\{\big[\bigotimes^{n}_{l=z+1}\big(\tilde{a}^{2}|0_{l}\rangle_{out}\langle 0_{l}|+\tilde{b}^{2}|1_{l}\rangle_{out}\langle1_{l}|\big)\big]\big[\bigotimes^{z}_{i=1}\big(\tilde{a}^{2}|0_{i}\rangle_{in}\langle 0_{i}|+\tilde{b}^{2}|1_{i}\rangle_{in}\langle 1_{i}|\big)\big]\big\}\bigg\}.
\end{eqnarray}

\begin{acknowledgments}
This work is supported by the National Natural
Science Foundation of China (Grant nos. 12175095, 12575056,  and 12205133 ),   LiaoNing Revitalization
Talents Program (XLYC2007047), and the
Special Fund for Basic Scientific Research of Provincial Universities in
Liaoning under Grant no. LS2024Q002.
\end{acknowledgments}


\end{document}